\documentclass[12pt,tightenlines,eqsecnum,floats,shownopacs,nofootinbib,amsmath,amssymb,aps,prd]{revtex4}

\usepackage{color}
\def\be{\nopagebreak[3]\begin{equation}}
\def\ee{\end{equation}}
\def\ba{\nopagebreak[3]\begin{eqnarray}}
\def\ea{\end{eqnarray}}

\def\lp{\ell_{\rm Pl}}

\def\f{\frac}
\def\mpl{m_{\rm Pl}}

\def\hom{\rm hom}
\def\inhom{\rm inh}
\def\dyn{\rm Dyn}
\def\cl{\rm cl}

\def\ub{\underbar}

\def\t{\tilde}
\def\h{\hat}
\def\sint{\textstyle{\int}}

\def\tr{\rm Trun}
\def\inter{\rm Int}

\def\pphi{p_{(\phi)}}

\def\ep{\epsilon}
\def\x{\vec{x}}
\def\vp{\varphi}
\def\vk{\vec{k}}

\def\vpone{\varphi^{(1)}}
\def\vptwo{\varphi^{(2)}}
\def\vpn{\varphi^{(n)}}
\def\vpthree{\varphi^{(3)}}
\def\pione{{\pi}^{(1)}}

\def\pin{{\pi}^{(n)}}
\def\C{\mathcal{C}}
\def\R{\mathcal{R}}
\def\Q{\mathcal{Q}}

\def\man{\mathbb{M}}
\def\ezero{\mathring{e}}
\def\ozero{\mathring{\omega}}
\def\qzero{\mathring{q}}
\def\Vzero{\mathring{V}}
\def\vzero{\mathring{v}}

\def\e{\mathfrak{e}}
\def\eone{{\mathfrak{e}^{(1)}}{}}
\def\etwo{{\mathfrak{e}^{(2)}}{}}
\def\en{{\mathfrak{e}^{(n)}}{}}
\def\a{\mathfrak{a}}
\def\aone{{\mathfrak{a}^{(1)}}{}}
\def\atwo{{\mathfrak{a}^{(2)}}{}}
\def\an{{\mathfrak{a}^{(n)}}{}}
\def\q{\mathfrak{q}}
\def\K{\mathfrak{K}}
\def\p{\mathfrak{p}}
\def\W{\mathfrak{W}}

\def\S{\mathcal{S}}
\def\emb{\mathcal{E}}
\def\T{\mathcal{T}}
\def\ps{\Gamma}
\def\Hkin{\mathcal{H}_{\rm kin}}
\def\H{\mathcal{H}}

\def\Hpo{\mathcal{H}_{\rm phy}^o}

\def\b{{\rm b}}
\def\B{{}_{\rm B}}%used for the bounce point
\def\w{\omega}

\begin{document}

\title{An Extension of the Quantum Theory of Cosmological Perturbations to the
Planck Era}

\author{Ivan Agullo${}^{1,2}$}
\email{I.AgulloRodenas@damtp.cam.ac.uk}
\author{Abhay Ashtekar${}^1$}
\email{ashtekar@gravity.psu.edu}
\author{William Nelson${}^1$}
\email{nelson@gravity.psu.edu} \affiliation{${}^1$ Institute for
Gravitation and the Cosmos \& Physics Department, Penn State,
University Park, PA 16802, U.S.A.\\
${}^2$ Center for Theoretical Cosmology, DAMTP, Wilberforce Road,
University of Cambridge, Cambridge CB3 OWA, U.K. }

\begin{abstract}

Cosmological perturbations are generally described by quantum fields
on (curved but) classical space-times. While this strategy has a
large domain of validity, it can not be justified in the quantum
gravity era where curvature and matter densities are of Planck
scale. Using techniques from loop quantum gravity, the standard
theory of cosmological perturbations is extended to overcome this
limitation. The new framework sharpens conceptual issues by
distinguishing between the true and apparent trans-Planckian
difficulties and provides sufficient conditions under which the true
difficulties can be overcome within a quantum gravity theory. In a
companion paper, this framework is applied to the standard
inflationary model, with interesting implications to theory as well
as observations.

\end{abstract}

\pacs{98.80.Qc, 04.60.Pp, 04.60.Kz}

\maketitle

\section{Introduction}
\label{s1}

This is the second in a series of papers whose goal is to
investigate whether current scenarios of the early universe admit
quantum gravity completions and, if they do, to study the
implications of the resulting Planck scale dynamics. The first paper
\cite{aan1} summarized the underlying framework and its applications
to inflation for a broad audience of theoretical physicists. This
paper is primarily addressed to the quantum gravity community and
provides a detailed extension of the cosmological perturbation
theory to the Planck regime. Specifically, we will consider gravity
coupled to a scalar field and study the dynamics of quantum fields
representing scalar and tensor perturbations on \emph{quantum}
cosmological space-times. In the third paper \cite{aan3}, addressed
to cosmologists, this framework is used to show that the
inflationary scenario admits a quantum gravity extension and to
analyze the physical implications of pre-inflationary dynamics, i.e.
the quantum evolution from the big bounce of loop quantum cosmology
(LQC) to the onset of the standard slow roll inflation. In the
future, we hope to examine whether alternatives to inflation also
admit viable quantum gravity completions and, if so, explore the
resulting Planck scale physics.

In the theoretical explorations of the early universe, one generally
uses the Friedmann, Lema\^itre, Robertson, Walker (FLRW) solutions
to the Einstein equations (with appropriate matter sources) as
background space-times. The focus is on the dynamics of quantized
fields representing linearized perturbations propagating on these
backgrounds. (See, e.g., \cite{ll-book,sd-book,vm-book,sw-book}.)
The necessary framework of quantum theory of linear fields in curved
space-times has been well developed, thanks to the ongoing research
that date back to the mid 1960s. (See, e.g.,
\cite{bd-book,sf-book,parker-book}). However, the FLRW space-times
of interest are invariably incomplete in the past due to the big
bang singularity where matter fields and space-time curvature
diverge. It is widely believed that general relativity is simply not
applicable once curvature reaches the Planck scale, whence there is
no justification for using quantum field theory on solutions to
Einstein's equations in this domain. Quantum gravity must intervene
in an important fashion. Thus, to encompass the Planck regime, one
needs a quantum gravity extension of the standard cosmological
perturbation theory.

Loop quantum gravity (LQG) provides a promising avenue to meet this
goal because by now the big bang singularity has been resolved in a
variety of models in LQC; the k=0 FLRW model on which we will focus
\cite{mb1,aps1,aps2,aps3,acs,ps}, as well as their generalizations
that include spatial curvature \cite{apsv,warsaw1}, a cosmological
constant \cite{bp,kp1,ap}, anisotropies
\cite{awe2,madrid-bianchi,awe3,we} and the simplest type of
inhomogeneities and gravitational waves
\cite{hybrid1,hybrid2,hybrid3,hybrid4,hybrid5}. (See, e.g.,
\cite{mbrev,asrev} for summaries of these developments.) It is
therefore natural to use LQC as the point of departure for extending
the cosmological perturbation theory. However, to do so, we cannot
just mimic the standard procedure used in general relativity because
LQG does yet offer the quantum version of full Einstein's equations
which one can linearize around a quantum FLRW space-time. Therefore
we will use the general strategy that has been repeatedly followed
in LQG: \emph{First truncate the classical theory in a manner
appropriate to the physical problem under consideration, then carry
out quantization using LQG techniques, i.e., paying due attention to
the underlying quantum geometry, and finally work out the physical
implications of the framework}. This strategy has led to advances in
the cosmological models referred to above, as well as in the
investigation of quantum black holes \cite{abck,abk,blv} and the
spin foam derivation of the graviton propagator
\cite{graviton1,graviton2,cr-rev}.

To extend the cosmological perturbation theory, then, we will begin
by focusing on the following sector of the full phase space of
general relativity: homogeneous, isotropic configurations
\emph{together with} first order inhomogeneous perturbations.
However, to encompass the Planck regime, we must now use a
\emph{quantum} FLRW geometry as background, and study the dynamics
of quantum fields representing scalar and tensor modes propagating
on these quantum geometries. Since such a quantum geometry provides
only probability amplitudes for various FLRW metrics, we no longer
have a sharply defined, proper or conformal time. How can one then
describe the dynamics of inhomogeneous perturbations? In \cite{akl},
this issue was resolved for test quantum fields on quantum FLRW
space-times by deparameterizing the Hamiltonian constraint in the
background, homogeneous sector. Then one can regard the background
scalar field $\phi$ as a relational time variable with respect to
which physical observables evolve. This is a new conceptual element,
made necessary by quantum gravity considerations. We will use the
same strategy.

However, to encompass cosmological perturbations, we will need three
significant extensions of Ref. \cite{akl}. First, while that work
discussed a test quantum scalar field, now the test fields include
metric perturbations. Second, to systematically arrive at the
evolution equation of perturbations on quantum geometry, one needs
an improved strategy. The third and most important difference is
that, because of its focus on conceptual issues such as the problem
of time, the analysis in \cite{akl} was restricted only to a finite
number of modes of the test scalar field, and thus avoided the
ultraviolet difficulties from the start. In this paper, by contrast,
we do not truncate the number of modes and much of the difficult
analysis is devoted to these ultra-violet issues.

Indeed, these issues play a central role in testing self-consistency
of our procedure. The key approximation underlying our truncation
strategy is that inhomogeneities can be regarded as perturbations
---i.e., their back-reaction on the space-time geometry can be
ignored. In the classical theory, a solution obtained using this
approximation is regarded as self-consistent if the stress-energy in
its inhomogeneities is negligible compared to that in the
homogeneous background for the entire dynamical regime of interest.
In our quantum theory of the truncated phase space, the situation
will turn out be the following. Fix a quantum FLRW background
geometry, described by a state $\Psi_o$, which evolves (with respect
to the deparameterized, internal time) via a Hamiltonian $\h{H}_o$.
The state undergoes a quantum bounce at some time $\phi=\phi_{\B}$.
There are natural initial conditions for quantum states $\psi$ of
perturbations at $\phi=\phi_{\B}$. We will show that for these
states, it suffices to focus just on the energy density, rather than
the full stress-energy. If the inhomogeneities are to be regarded as
perturbations, we have to choose the initial $\psi$ so that the
energy density in it is small at $\phi=\phi_{\B}$. The question is
whether it continues to remain small in the entire duration of
evolution of interest. This is not at all guaranteed, especially
because of the Planck scale curvature during and following the
superinflation phase immediately after the bounce. But if it does,
then $\Psi_o\otimes\psi$ would be a self-consistent solution to the
truncated quantum theory.%
\footnote{Of course, self-consistency by itself does not imply that
a truncated solution is necessarily close to an exact one. This is
so even in classical general relativity because, even if the first
order perturbation remains very small, the sum of all higher order
terms could be large. Yet, in classical cosmology and investigations
of black hole perturbations, if the back-reaction due to first order
perturbations remains negligible ---i.e., if the test field
approximation is self consistent--- the first order truncation is
generally regarded as a good approximation. We adopt the same
viewpoint in this paper.}
Note that the very formulation and subsequent analysis of this
self-consistency criterion requires a well-controlled definition of
the Hamiltonians $\h{H}_o$ and $\h{H}_1$ (and the corresponding
energy density operators). Existing results on LQC already provide a
well-defined, specific $\h{H}_o$. On the other hand, $\h{H}_1$ is a
\emph{composite} operator on the Hilbert space of perturbations; its
formal expression involves products of operator valued
distributions. Therefore, an appropriate ultraviolet regularization
and renormalization is essential, first to obtain a well-defined
evolution of $\psi$, and second to check that the energy density in
the state $\psi$ continues to remain small compared to that in the
background. To properly handle this issue, we will carry over the
well developed techniques of adiabatic regularization from the
quantum field theory on classical FLRW space-times to that on
quantum FLRW space-times.

Because this article is addressed primarily to the quantum gravity
audience, we will make two assumptions that will enable us to have
good mathematical control without unduly simplifying the essential
conceptual underpinning. First, we will assume that the spatial
topology is that of a 3-torus $\mathbb{T}^3$ rather than
$\mathbb{R}^3$. (In practice, there would be no obvious conflicts
with CMB observations if one lets the physical radius of each of the
three $S^1$ in $\mathbb{T}^3$ at the last scattering surface to be
greater than the known radius $R_{\rm obs}$ of the observable
universe at that time \cite{torus}.) This assumption lets us
cleanly avoid spurious infinities that would arise with an
$\mathbb{R}^3$ topology simply because the background fields are
homogeneous. However, our main results extend to the $\mathbb{R}^3$
topology and at various junctures we will indicate how these issues
are handled in that case. Second, we will assume that there is no
potential; $V(\phi)=0$. To incorporate inflation, one has to remove
this assumption and the necessary modifications are summarized in
the Appendix A of Ref. \cite{aan3} which discusses inflation in
detail.

The paper is organized as follows. To put this work in a broader
perspective, in section \ref{s2} we first summarize a few procedures
that have been followed in the literature to extend the cosmological
perturbation theory using LQC, point out their merits and
limitations, and then present our strategy. In section \ref{s3} we
spell out the desired truncation in the classical theory.
Specifically, we start with the full phase space of general
relativity coupled to a scalar field, truncate the constraints to
second order in inhomogeneities, and express the Hamiltonian $H_1$
governing dynamics of the first order perturbations in terms of
those gauge invariant variables that are most convenient for passage
to the quantum theory. In this passage, certain conceptual
subtleties arise that affect subsequent technical results. We
discuss them in some detail because, while they are well known in
the discussion of perturbations of black holes, they are often
overlooked in the cosmological context, especially in the LQG
literature. (See Appendix A for a discussion of these issues in a
simpler context of the $\lambda \phi^4$ theory in Minkowski
space-time.)

In sections \ref{s4} - \ref{s6} we discuss the quantum theory. The
main steps involved in the construction are first collected in
section \ref{s4}; the detailed construction of the Hilbert space of
states follows in \ref{s5}; and the necessary regularization and
renormalization of composite operators is then discussed in section
\ref{s6}.

Physical states $\Psi(\nu,\phi; \Q, \T,)$ depend on $\nu \sim a^3$
(the cube of the scale factor), the background scalar field $\phi$,
and on the gauge invariant scalar mode $\Q$ and the two tensor modes
$\T$ of inhomogeneous perturbations. As is standard in LQC, the form
of (the homogeneous part of the) quantum constraint will enable us
to regard $\phi$ as a relational time variable so that, at a
fundamental level, dynamics refers to this internal or emergent
time. Since the back-reaction of perturbations is neglected, one can
express $\Psi(\nu,\phi; \Q, \T)$ as a tensor product $\Psi_o (\nu,
\phi)\otimes \psi (\Q, \T, \phi)$ and study its evolution. Then, a
surprising and key simplification occurs: The evolution of the
quantum fields $\h{\Q}, \h\T$ on the \emph{quantum} geometry defined
by $\Psi_o$ (in relational time $\phi$) is mathematically equivalent
to their evolution on an effective background metric $\t{g}_{ab}$
(in its conformal time $\t\eta$) where $\t{g}_{ab}$ is an effective
metric, `dressed' with quantum corrections. Physically, while the
evolution of the quantum fields $\h{\Q}, \h\T$ \emph{is} sensitive
to the quantum nature of geometry defined by $\Psi_o$, it does not
`see' all the details of this quantum geometry: It is sensitive only
to the expectation value and certain aspects of the fluctuations of
the quantum metric which, it turns out, can be captured by
$\t{g}_{ab}$. At late times $\t{g}_{ab}$ can be well approximated by
a FLRW solution to Einstein's equation. Therefore, this passage from
the relational time $\phi$ to $\t\eta$ also makes it manifest that,
while the quantum constraint provides an evolution w.r.t. $\phi$
starting right from the bounce, away from the Planck regime this
evolution reduces to that used in the familiar treatment
\cite{ll-book,sd-book,vm-book,sw-book} of quantum perturbations.

The key question, as we have already indicated, is whether our basic
assumption that the back-reaction can be neglected is borne out in
the resulting solutions, especially in the Planck era. The technical
discussion of section \ref{s6} is devoted to this issue and leads to
a sufficient condition for the solution $\Psi_o (\nu, \phi)\otimes
\psi (\Q, \T, \phi)$ to be self-consistent within the truncation
approximation. In the next paper \cite{aan3} we will show that for a
large class of initial conditions, this criterion is met in presence
of a quadratic potential $V(\phi)$ all the way from the bounce to
the onset of slow roll inflation. In section \ref{s7} we summarize
the main results and discuss various issues, including the choice of
initial conditions.

Our conventions are as follows. The space-time metric will be
assumed to have signature -,+,+,+. we set $c=1$ but keep ($G$ and)
$\hbar$ explicitly in various equations to facilitate the
distinction between classical and quantum effects. Finally, as in
the quantum gravity literature, we will use Planck (rather than the
reduced Planck) units. (Thus, our $\lp = \sqrt{\hbar G}$ and our
Planck mass $\mpl= \sqrt{\hbar/G}$ is related to the reduced planck
mass $M_{\rm Pl}$ via $\mpl = \sqrt{8\pi} M_{\rm Pl}$.)

\section{Strategies inspired by LQC}
\label{s2}

The singularity resolution in LQC has motivated a large number of
investigations aimed at incorporating the underlying quantum
geometry effects in the standard cosmological paradigms. In this
section we will first briefly summarize the main strategies and then
explain the approach that is followed in the rest of the paper. This
concise summary should help non-experts to see the inter-relation
(or lack thereof) between the rich set of ideas that are being
pursued in the LQG literature. For experts, it should help clarify
some subtle issues that have not been emphasized in the literature
and also bring out the continuity and coherence of the approach used
here.

\subsection{A brief summary}
\label{s2.1}

Recall first that in the FLRW models the LQC quantum states of
interest remain sharply peaked along the bouncing solution of
effective equations that incorporate leading order quantum
corrections. This surprising behavior was first encountered in
numerical simulations but subsequent work led to an analytic
understanding through a change of representation \cite{acs}, use of
coherent states \cite{vt} and the WKB approximation \cite{ach3}.
Early attempts sought to exploit this property to incorporate
quantum gravity corrections right before the onset of the slow roll
inflation \cite{tsm}, or, for the evolution of perturbations during
inflation (see, e.g., \cite{als}). However, to incorporate these
corrections to the evolution, the LQC effective equations for the
FLRW background do not suffice; one also needs the LQG modified
perturbation equations. Since a well established set of quantum
Einstein's equations is not yet available in full LQG, the strategy
was to simply use the first order perturbation equations from
general relativity, the FLRW background space-time being simply
replaced by the effective solution in LQC. Unfortunately,  this
procedure is conceptually unsatisfactory because the background LQC
space-time is no longer a solution to Einstein's equations. Indeed,
now there is an ambiguity in what one means by `linearized
Einstein's equations': Two sets which are equivalent when the
background is a solution to the exact Einstein's equations are
generically no longer equivalent if the background is a solution to
a different set of equations. To our knowledge, a systematic
procedure to handle this ambiguity was not part of the general
strategy.

Another set of papers, geared to capture the `inverse triad (or
inverse volume) corrections'  of LQC, is based on the notion of
`lattice refinement' (see, e.g., \cite{ns_inflation,bct,gc}). In the
homogeneous, spatially compact case, these corrections are
meaningful and physically interesting. In the spatially non-compact
case, typically the corrections depend on the choice of a fiducial
cell used as an infrared regulator \cite{aps3} and disappear when
the regulator is removed \cite{asrev}. In more recent versions, one
considers $\mathbb{R}^3$ spatial topology but decomposes the spatial
manifold in elementary cells and approximates the inhomogeneous
configurations of physical interest by configurations which are
homogeneous within any one cell but vary from one cell to another.
Physically, this is an attractive strategy. However it requires a
fresh input ---that of the cell size (or `lattice spacing')--- and
the inverse volume effects are now sensitive to this new scale: The
fiducial cell of the homogeneous model is in effect replaced by a
physical cell in which the universe can be taken to be homogeneous.
However, so far there is neither a theoretical principle nor
observational guidance on what this scale should be during the
inflationary epoch when perturbations are generated in these
schemes. On the other hand, the more striking predictions of these
frameworks ---such as enhancement of quantum gravity signature by
several orders of magnitude over the factor $H^2/\mpl^2 \sim
10^{-11}$ one would expect during inflation--- appear to depend on
the choice of the new scale one makes. Therefore, although the
underlying idea of lattice refinement is attractive, at present
there appears to be an inherent ambiguity in the size and importance
of the inverse volume effects in this setting. Overcoming these
limitations is an interesting prospect for future work.

In these investigations the focus was on quantum gravity effects
during inflation. Some of the more recent investigations have
recognized that, because the energy density is $\sim 10^{-11}
\rho_{\rm Pl}$ during this era, quantum gravity corrections during
inflation would be too small to be observable in the foreseeable
future and shifted the emphasis to studying quantum gravity
corrections from the bounce to the onset of the slow roll. Many of
them employ a new strategy that goes under the broad theme of
`anomaly cancelation'
\cite{pert_tensor1,pert_anomaly,barrau1,barrau2,barrau3,barrau4,barrau5}.
This analysis is based on the Hamiltonian framework and focuses on
the constraint algebra. The idea is to arrive at the desired LQG
theory of cosmological perturbations by studying the permissible
modifications of the constraints of general relativity that are to
encode quantum corrections. Recall that two key features of general
relativity are: i) the Poisson algebra of constraints closes; and,
ii) the evolution is generated by these constraints. Keeping these
features in mind, one proceeds in the following steps: i) One
assumes that the `effective theory' that incorporates the LQG
corrections would have the same phase space for the homogeneous
isotropic background and first order perturbations; ii) allows for a
modified set of constraints on this phase space by making an ansatz
for possible modifications; iii) calculates the undetermined
functions of the background geometry in the ansatz by
\emph{requiring} that the constraint algebra should again close;
and, finally, iv) defines the desired effective dynamics as the
Hamiltonian flow of the new, modified constraints. In the quantum
theory, if the commutator algebra of constraints did not close,
there would be an anomaly. Therefore, the method is called `anomaly
cancelation' even though one is dealing only with Poisson brackets.

Until recently, the modifications that were arrived at did not
change the structure functions in the constraint algebra of general
relativity. In a recent work on scalar perturbations \cite{barrau5},
on the other hand, the structure function in the Poisson bracket
between two Hamiltonian constraint is modified by a function of the
ratio $\rho/2\rho_{\max}$  where $\rho$ is the matter density in the
background and $\rho_{\max}$ the maximum density in FLRW LQC. This
has been interpreted to mean that there is a change in the
space-time signature at the end of the superinflation phase of the
background, so that the signature is Euclidean to the past of this
event \cite{mb-gp}.

The central idea in this `anomaly cancelation' strategy is
potentially powerful in constraining the type of quantum corrections
cosmological perturbation theory can inherit from \emph{any}
consistent quantum gravity theory. However, its implementation has
some puzzling aspects.% and undesirable features.

First, the form of permissible modifications of the constraint
functionals is chosen primarily for mathematical convenience and not
derived systematically from general physical principles. Second, the
phase space of the quantum corrected theory is assumed to be the
same as that in classical general relativity while typically,
quantum corrections add higher derivative terms to the action which
significantly enlarge the phase space. Third, since the phase space
is kept unchanged, the general analysis due to Hojman, Kuchar and
Teitelboim \cite{hkt} is applicable and their results bring out a
puzzling feature of this framework. Hojman et al began with
4-dimensional, globally hyperbolic space-times $(M,g_{ab})$ with
non-spinorial matter and, using embeddings of a 3-manifold $\man$ as
a Cauchy surface in $M$, constructed a `hypersurface deformation'
algebra $\mathcal {D}$ on the space $\emb$ of embeddings which
encodes space-time covariance. This is a purely geometric
construction without any reference to field equations. Then they
showed that, on the standard phase space of general relativity
(coupled to the matter under consideration) based on $\man$, there
is only one way to represent the algebra $\mathcal{D}$ by canonical
transformations in a time reversible manner: the representation
given by the Poisson algebra of the \emph{standard} Einstein
constraint functionals. For the anomaly cancelation program, this
implies that if the modification of constraint functionals is
genuine (i.e., not just a field re-definition), then the modified
Hamiltonian theory will not have a consistent space-time
interpretation. Therefore would not be possible to associate a
well-defined space-time metric to (portions of) dynamical
trajectories in this modified Hamiltonian theory, let alone examine
its signature. Fourth, even if one were to ignore this point,
signature change is such a drastic effect that it seems difficult to
justify the validity of the first order cosmological perturbation
theory in the subsequent treatment. The fifth and a more `global'
limitation arises from the fact that the conceptual underpinning of
the overall strategy is rather unclear in some of the recent
applications. Effective equations are meant to incorporate all
quantum corrections. Therefore, one would have thought that the
dynamical equations derived from them would already contain quantum
effects. Yet, in some works, these fields are quantized again to
obtain the power spectrum for scalar, vector and tensor modes, as in
the standard treatment of cosmological perturbations on classical
FLRW backgrounds in general relativity. The overall logic underlying
this scheme is thus rather puzzling. Indeed, already in the
homogeneous sector the logical procedure is the opposite: one first
obtains the quantum evolution equations and \emph{then} derives
effective equations from them using dynamics of appropriate, sharply
peaked coherent states.

To summarize, investigations to date have provided useful
mathematical infrastructure (see, e.g., \cite{pert_tensor1}) and in
some cases also the much needed qualitative insights into mechanisms
through which quantum gravity effects could provide corrections to
the standard inflationary scenario (see, e.g.,
\cite{jm,barrau3,barrau4,barrau5}). The viewpoint that appears to
have emerged from the `anomaly cancelation' strategy
---namely, quantum effects could be neatly encoded in the Hamiltonian
framework but would lead to a fuzziness if the phase space
trajectories are interpreted as 4-dimensional space-time
geometries--- could well be an imprint of some deep result on the
nature of space-time in LQG. However, it seems fair to say that, at
the current level of understanding, this and other strategies used
so far also have a number of puzzling features and face conceptual
limitations in their treatment of inhomogeneous perturbations.

% IN THE CONCLUSION WE SHOULD DISCUSS OTHER LIMITATIONS/PUZZLING FEATURES:
%vector perturbations; critical density disappearing from tensor mode
%equations; no reference to Hadamard; no control of back-reaction; No
%reference to WMAP data; Quite a bit of analysis irrelevant for
%observable modes;

\subsection{Strategy used in this paper}
\label{s2.2}

In this sub-section we will outline the avenue used in this paper to
extend the standard theory of cosmological perturbations all the way
to the Planck regime. As explained in section \ref{s1}, the main
idea is simply to use the strategy that has driven LQG so far:
Construct the Hamiltonian framework of the sector of general
relativity of interest and then pass to the quantum theory using
quantum geometry that underlies LQG. In order to bring out
differences from other approaches, and to address questions
regarding truncation and gauge choices that are sometimes raised, we
will now spell out this strategy using illustrations where is has
already been successfully used.

The FLRW models provide the simplest illustration. Here, one starts
out with the full phase space $\ps$ of general relativity (coupled
to matter) in the connection dynamics framework \cite{aa-newvar,fb}
and truncates it to its homogeneous, isotropic sector $\ps_{\rm
HI}$. By fixing gauge, one coordinatizes the gravitational part of
$\ps_{\rm HI}$ simply with a pair $(\nu, \b)$ of real numbers, where
(in appropriate units) $\nu$ denotes the physical volume of the
universe (or, if the spatial topology is non-compact, of a fiducial
cell) and $\b$ is its conjugate momentum. This gauge fixing leads us
to the \emph{classically reduced phase space} with respect to the
Gauss and the diffeomorphism constraints. Therefore the reduced
phase space carries only the Hamiltonian constraint, expressed in
terms of just $\nu, \b$ and matter variables. This provides the
starting point for quantization a la LQG. In the first step, a
specific kinetic framework \cite{abl} can be selected by a
uniqueness theorem \cite{ac} along the lines of those in full LQG
\cite{lost,cf}. By representing curvature in the Hamiltonian
constraint by holonomies around plaquettes selected by the
underlying quantum geometry, one constructs the quantum Hamiltonian
constraint operator \cite{aps3,asrev}. In this construction there
are, as is usual in any quantization procedure, some factor ordering
ambiguities. (Compare, e.g., the Hamiltonian constraint in
\cite{aps3} (which is geared to proper time) with that in \cite{acs}
(geared to harmonic time). See also \cite{sw}.) However, these
affect only details of the quantum evolution; general features are
robust. Finally, the effective equations that encode leading quantum
effects are systematically derived starting from the quantum theory
\cite{vt}.

In the generalization to the homogeneous but anisotropic (i.e.,
Bianchi) models, one again follows the same conceptual procedure:
appropriate truncation of the phase space $\ps$ to $\ps_{\hom}$,
eliminating the Gauss and diffeomorphism constraints by passing to
the \emph{classically reduced phase space}, and passing to quantum
theory via LQG techniques \cite{awe2,awe3,we}. The singularity
resolution persists. Furthermore, there is a detailed consistency
check: by tracing over the anisotropy degrees of freedom of the
Bianchi I model one recovers the quantum Hamiltonian constraint of
the FLRW model \cite{awe2}. As one would expect, in presence of
anisotropies, the quantum dynamics is significantly richer
\cite{gs,ps-bianchi,ac-bianchi}. Note that the starting point in
this analysis is again a truncated phase space $\ps_{\hom}$ of
general relativity, but the truncation is now enlarged to
incorporate physics of interest, namely anisotropies.

The last example, Gowdy models,
\cite{hybrid1,hybrid2,hybrid3,hybrid4,hybrid5} illustrate our
strategy most closely because now the truncated phase space
$\ps_{\rm Gowdy}$ is an \emph{infinite} dimensional subspace of
$\ps$: the model allows for inhomogeneities induced by a class of
(fully non-linear) gravitational waves, in addition to those in the
matter fields. The constraints are now rewritten in terms of an
infinite number of conveniently chosen `modes' of canonical
variables coordinatizing $\ps_{\rm Gowdy}$, obtained by appropriate
gauge fixing. Thanks to this choice, we are led to a \emph{classical
reduction of the phase space} with respect to the Gauss and the
purely inhomogeneous parts of the diffeomorphism and Hamiltonian
constraints. Thus, one is left with only  `global' Hamiltonian and
diffeomorphism constraints (corresponding to a homogeneous lapse and
shift). One then passes to quantum theory using a `hybrid' scheme
where one employs LQC quantum kinematics for homogeneous modes and a
Fock-type quantum kinematics for the inhomogeneous modes
representing gravitational waves. This is an internally consistent
quantization. One finds that, thanks to the quantum geometry
effects, the singularity of general relativity is resolved. This is
in striking contrast to the early attempts predating LQC, where
singularity could not be resolved (see, e.g.
\cite{gowdy-old1,gowdy-old2,gowdy-old3,gowdy-old4}). It is
interesting to note that, if one `switches off' inhomogeneities,
Gowdy models reduce to Bianchi I models. Using effective equations,
it has been shown that the general dynamical behavior of the Bianchi
I models ---including the bounces--- carries over to the Gowdy
models \cite{hybrid2,hybrid5}. This analysis has also provided
valuable information on the changes in the amplitudes of
gravitational waves resulting from the bounce. Finally note that,
although one obtains the Bianchi I model by switching off
inhomogeneities, in contrast to some of the strategies summarized in
section \ref{s2.1}, one \emph{does not} begin with the inhomogeneous
modes propagating on an effective, quantum corrected Bianchi I
background and \emph{then} quantize them. Rather, one truncates the
full phase space $\ps$ to $\ps_{\rm Gowdy}$ and \emph{quantizes the
full Gowdy model, which includes both homogeneous and inhomogeneous
modes.}

To develop a quantum gravity theory of cosmological perturbations,
then, we will continue along the same path that has been
successfully used so far. Our first task is to identify the
appropriate truncation of the classical phase space $\ps$. By
introducing a fiducial flat triad for mathematical convenience, we
can decompose fields into Fourier modes.  With this convenient
coordinatization, constraints of the full theory can be expressed in
terms of modes. We will start with a homogeneous background and
expand out the deviations from it as a sum of first, second, ...
nth, ... order terms in inhomogeneities. The truncation will now
consist of keeping terms which depend on the background and are at
most quadratic in the first order perturbations. In general
relativity, this truncation enables one to study dynamics of the
background homogeneous space-time and that of linearized
perturbations propagating on these backgrounds. Back-reaction of
these first order perturbations on the background (which is encoded
in the second
order perturbations) is neglected.%
\footnote{There are some subtleties in this procedure that are
sometimes overlooked. For a discussion in a simpler example, see
Appendix \ref{a1}.}

The idea again is to use LQG techniques to pass to the quantum
theory, using a hybrid scheme that broadly mimics the one used in
the Gowdy models. (As we will see in subsequent sections, some
differences arise because, unlike in the Gowdy model, we are now
dealing with \emph{linear perturbations}.) It will again be possible
to interpret the homogeneous quantum Hamiltonian constraint as
providing `evolution' of the quantum state of the background
geometry in the relational time variable ---the scalar field.
Solutions to this equation provide background \emph{quantum}
geometries on which quantum perturbations evolve. In this conceptual
setting, one does \emph{not} start out with classical perturbations
on an effective, smooth FLRW space-time and \emph{then} quantize
them. The passage to quantum theory is carried out in one go for the
full truncated phase space that includes both the background and
perturbations. Finally, as emphasized in section \ref{s1}, much of
the technical discussion will be devoted to finding a criterion to
test whether the final theory admits self-consistent solutions that
justify the viability of the underlying truncation scheme.

This overall strategy was briefly reported in \cite{ia-rep,wn-rep}.
At the same time the `hybrid approach' was used in \cite{madrid1} to
study cosmological perturbations in the $k$=1 FLRW context. The main
conceptual difference between that approach and ours is that we are
able to go beyond formal considerations in the quantum theory by
exploiting the relation between quantum fields on quantum geometry
and those on a dressed effective geometry discussed in section
\ref{s4.2}.

\section{Truncated Hamiltonian framework}
\label{s3}

In cosmology of the very early universe, one generally restricts
oneself to the sector of general relativity consisting of
homogeneous FLRW space-times together with linear perturbations
thereon. In much of the cosmology literature one works with the
\emph{solution space} of this truncated theory. However, as pointed
out in Ref. \cite{langlois}, the task of finding gauge invariant
variables is more stream-lined in the Hamiltonian framework. More
importantly, since we are now interested in treating the background
geometry quantum mechanically, a natural avenue is to follow the
Dirac quantization procedure based on phase space. Therefore, in
this section we will first construct the truncated phase space and
then discuss dynamics thereon. This will provide a natural starting
point to apply the well established LQG techniques in the next
section.

\subsection{The Phase space}
\label{s3.1}

Let us begin with general relativity coupled to a scalar field on a
space-time manifold $M= \man \times \mathbb{R}$, where $\man$ is
topologically $\mathbb{T}^3$. For completeness and continuity with
the LQC literature, we will begin with the connection variables
\cite{aa-newvar} and then pass to the Arnowitt Deser Misner (ADM)
variables for perturbations that are more commonly used in the
cosmological literature.

Let us first focus on geometry. Let $q_{ab}$ denote positive
definite metrics on $\man$, $e^a_i$ and $\omega_a^i$, orthonormal
frames and co-frames with respect to $q_{ab}$, and let $K_{ab}$
denote the extrinsic curvature on $\man$. In connection dynamics,
the canonically conjugate pair consists of real ${\rm SU(2)}$
connections $A_a^i$ and ${\rm su(2)}$ valued vector densities
$E^a_i$ of weight 1, both defined on $\man$. (Here indices $a,b,c,
\ldots$ refer to the tangent space of $\man$ and $i,j, k, \ldots$ to
the Lie algebra ${\rm su(2)})$. They are related to the ADM
variables via:
\be \label{relation} E^a_i = \sqrt{q}\, e^a_i \quad {\rm and} \quad
A_a^i = \Gamma_a^i + \gamma K_a^i \ee
where $q$ denotes the determinant of $q_{ab}$, \, $\Gamma_a^i$, the
spin connection determined by $e^a_i$, $K_a^i = K_{ab}e^{bi}$ and
$\gamma$, the Barbero-Immirzi parameter of LQG. Since we are
interested in the spatially flat FLRW background geometries, as
usual it is convenient to introduce some fiducial structures. Fix a
flat metric $\qzero_{ab}$, an orthonormal frame $\ezero^a_i$, and
the corresponding co-frame $\ozero_a^i$ on $\man$. We will denote by
$\qzero$ the determinant of $\qzero_{ab}$ and assume that each of
the circles in $\man$ has length $\ell$ with respect to
$\qzero_{ab}$ so that the fiducial volume of $\man$ is $\Vzero =
\ell^3$. The natural `Cartesian' coordinates defined on $\man$ by
$\qzero_{ab}$ will be denote by $\x \equiv (x_1, x_2, x_3)$.

Even though we are still considering full general relativity, it is
convenient to decompose the basic canonically conjugate fields into
purely homogeneous and purely inhomogeneous parts:
\be \label{decompose1} A_a^i(\x) = c \ell^{-1} \ozero_a^i +
\a_a^i(\x) \quad {\rm and} \quad E^a_i(\x) = \sqrt{\qzero}\,(p
\ell^{-2} \ezero^a_i + \e^a_i (\x)) \ee
where $\int (A_a^i \ezero^a_j) d\vzero =: c \ell^2 \delta_i^j$ and
$\int (E^a_i \ozero_a^j) d^3\x =:  p \ell \delta_i^j$ so that
$\a_a^i$ and $\e^a_i$ %have a suitable fall-off and
are \emph{purely inhomogeneous}:
\be \label{inhom} \sint \a_a^i \ezero^a_j\, d\vzero =0, \quad\quad
{\rm and} \quad\quad \sint \e^a_i \ozero_a^j\, d\vzero = 0\, .\ee
(Here and in what follows, unless otherwise specified, the integrals
are over $\man$ and $d\vzero$ denotes the volume element on it with
respect to $\qzero_{ab}$.) Thus, the geometrical part of the phase
space is naturally coordinatized by quadruples $(c,p; \a_a^i(\x),
\e^a_i (\x))$ where $c,p$ are real numbers and $\a_a^i, \e^a_i$ are
purely inhomogeneous fields. The matter field can also be decomposed
into purely homogeneous and purely inhomogeneous parts:
\be \label{decompose2} \Phi(\x) = \phi + \vp(\x) \quad {\rm and}
\quad \Pi(\x) = \sqrt{\qzero}\, (\ell^{-3}\pphi +  \pi(\x))  \ee
The symplectic structure on the total phase space $\ps$ is given by
\ba \Omega(\delta_1, \delta_2) &=& \f{3}{\kappa\gamma} [\delta_1 c
\delta_2 p - \delta_2 c \delta_1 p]\, +\, \sint (\delta_1 \a_a^i
\delta_2 \e^a_i - \delta_2 \a_a^i \delta_1 \e^a_i)\, d\vzero \nonumber\\
&+& \delta_1 \phi \delta_2 \pphi - \delta_2 \phi \delta_1 \pphi
\,+\, \sint (\delta_1 \varphi \delta_2 \pi - \delta_2 \pi \delta_1
\pi)\, d\vzero \ea
where $\delta \equiv (\delta c, \delta p, \delta\a_a^i,
\delta\e^a_i; \delta \phi, \delta\pphi, \delta\vp, \delta\pi)$
denote tangent vectors to $\Gamma$ and $\kappa = 8\pi G$. Thus, the
only non-zero Poisson brackets between the basic variables are:
\ba \label{pbs}\{c,\, p\} = \f{\kappa\gamma}{3}, &\quad&
\{\a_a^i(\x_1),\, \e^b_j(\x_2)\} = \delta^i_j\, \delta_a^b\,
\bar\delta(\x_1,\x_2),\nonumber\\
\{\phi,\, \pphi\} = 1, &\quad& \{\vp(\x_1),\, \pi(\x_2)\} =
\bar\delta(\x_1, \x_2), \ea
where $\bar\delta(\x_1,\x_2) = ((1/\sqrt{\qzero})\delta(\x_1,\x_2)\,
- (1/\ell^3))$ is the Dirac delta distribution on the space of
purely inhomogeneous fields.%
\footnote{The $1/\ell^3$ factor ensures that the Poisson brackets
are compatible with the fact that the perturbations are purely
inhomogeneous, i.e., satisfy (\ref{inhom}).}
Thus, the total phase space $(\ps, \Omega)$ has a product structure:
$\ps = \ps_{\hom} \times \ps_{\inhom},\,\, \Omega = \Omega_{\hom} +
\Omega_{\inhom}$.

There are three sets of first class constraints:
\ba \mathbb{G}[\Lambda] &=& \sint \Lambda^i(\x) G_i(\x)\, d^3\x  =
\sint \Lambda^i(\x)\, D_aE^a_i (\x)\, d^3\x \nonumber\\
\mathbb{V}[\vec{N}] &=& \sint N^a(\x) V_a (\x)\, d^3\x, \qquad
\mathbb{S}[N] = \sint N(\x) S(\x)\, d^3\x \, .\ea
Here, the vector and the scalar constraints are smeared by a shift
$\vec{N}$ and a lapse $N$ and the Gauss constraint ---which we have
explicitly written out because it does not feature in the ADM
framework--- by a generator $\Lambda^i$ of ${\rm su(2)}$. %(As usual,
%all the smearing functions are assumed to fall off at infinity
%sufficiently fast.)
The Gauss and the vector constraints generate, respectively,
internal ${\rm SU(2)}$ rotations and spatial diffeomorphism, both of
which are generally regarded as `kinematic motions.' Dynamics is
generated by the Hamiltonian constraint.\\

\emph{Remark:} If the spatial topology is $\mathbb{R}^3$ rather than
$\mathbb{T}^3$, the symplectic structure on the homogeneous subspace
induced by that of full general relativity diverges. Therefore, to
obtain a consistent phase space description, one has to introduce a
cubical fiducial cell $\C$ aligned with these axes and with
edge-length $\ell$ and restrict integrations of homogeneous fields
to it. This is an infrared cut-off, to be removed, as usual, in the
\emph{final} results (see e.g. \cite{asrev}).

\subsection{Expansions around the FLRW subspace}
\label{s3.2}

In physical cosmology one is primarily interested in a neighborhood
of the 4-dimensional, homogeneous sub-space $\ps_{\hom}$ of the full
phase space $\ps$. To fix notation for the rest of the paper, we
begin with certain expansions of fields near homogeneity. While this
technique is well known in the perturbation theory around black
holes, it appears not to be as familiar in the cosmology literature.

Consider curves $\gamma[\ep]$ in $\ps$ parameterized by $\ep$, with
$\epsilon \in ]-1,\, 1[$, say, which pass through $\ps_{\hom}$ at
$\ep=0$:
\ba \label{expan} A_a^i[\ep](\x) &=& c \ell^{-1}\ozero_a^i + \ep\,
\aone_a^i(\x) + \ldots + \f{\ep^n}{n!}\, \an_a^i(\x) + \ldots
\nonumber\\
E^a_i[\ep](\x) &=& \sqrt{\qzero}\, [p \ell^{-2}\ezero^a_i + \ep\,
\eone{}^a_i(\x) +
\ldots + \f{\ep^n}{n!} \en{}^a_i(\x) + \ldots] \nonumber\\
\Phi[\ep](\x) &=& \phi + \ep\, \vpone(\x) + \ldots + \f{\ep^n}{n!}
\vpn (\x) + \ldots\nonumber\\
\Pi[\ep](\x) &=& \sqrt{\qzero}\, [\ell^{-3}\pphi + \ep\, \pione(\x)
+ \ldots + \f{\ep^n}{n!} \pin (\x) + \ldots ] \ea
where $\aone_a^i, \, \eone^a_i, \vpone, \pione$ are \emph{purely
inhomogeneous} tangent vectors to the curves $\gamma[\ep]$ pointing
away from $\ps_{\hom}$. Here $\ep$ is a mathematical parameter that
keeps track of the `order' of the perturbation (it could be taken to
be $\sqrt{\kappa}$ but for simplicity we will assume it to be
dimensionless here). Geometrically, keeping only the first order
perturbations corresponds to considering the normal bundle over the
homogeneous subspace $\ps_{\hom}$ of $\Gamma$ (since purely
inhomogeneous tangent vectors are `orthogonal to' $\ps_{\hom}$ in
the $L^2$-norm). Retaining terms only up to the $n$th order
corresponds to considering the $n$th order, inhomogeneous jet bundle
on $\ps_{\hom}$. (Appendix A summarizes the meaning and utility of
this expansion procedure using the simpler example of the
$\lambda\Phi^4$ theory.)

Of special interest are the curves $\gamma[\ep]$ that lie in the
constraint hypersurface of $\ps$. We will use a collective label
$C(A,E, \Phi,\Pi)$ for the smeared constraint functions (suppressing
the smearing fields for simplicity). Then along each curve
$\gamma[\ep]$ the constraints become $C[\ep] =0$. By Taylor
expanding in $\ep$ we obtain a hierarchy of equations:
\be C|_{\ep =0}= 0,\quad \f{d C}{d\ep}|_{\ep=0} = 0, \quad
\ldots\quad \f{d^n C}{d\ep^n}|_{\ep=0} = 0,\quad \ldots \ee
The zeroth-order equation in the hierarchy, $C|_{\ep=0} = 0$ is just
the restriction of the full constraint to the homogeneous subspace
$\ps_{\hom}$. As noted above, because of gauge fixing, the Gauss and
the vector constraints vanish identically and we are left with only
one non-trivial constraint smeared with a homogeneous lapse
\cite{asrev}:
\be \label{homscalar1}\mathbb{S}_o[N_{\hom}]\, := \, N_{\hom}\,
[-\f{3}{\kappa\gamma^2}\, c^2 |p|^{\f{1}{2}} + \f{\pphi^2}{2
|p|^{\f{3}{2}}} ] = 0\, . \ee
(As noted in section \ref{s1}, in this paper we have set the
potential $V(\Phi)$ of the scalar field to zero. See the Appendix in
\cite{aan3} for inclusion of the potential). As usual, although the
lapse is homogeneous, it can depend on dynamical variables; see
section \ref{s3.4} for a further discussion. Eq (\ref{homscalar1})
is a non-linear but \emph{algebraic} equation constraining the
homogeneous fields.

The first order equation, $[dC/d\ep]|_{\ep=0} =0$, is a linear
partial differential equation (PDE) for $\aone_a^i(\x),
\eone^a_i(\x)$, $\vpone(\x), \pione(\x)$. For example, the Gauss
constraint yields:
\be \label{gauss1} \sint \Lambda^b_{\inhom} \, [ \ell \partial^a
\eone_{ab} + c\, \mathring{\ep}_b{}^{mn}\, \eone_{mn} - p\,
\ell^{-1}\, \mathring{\ep}_b{}^{mn}\aone_{mn}]\, d{\vzero}\, = 0 \ee
where we have converted the internal indices $i,j,\ldots$ into
tangent space indices $a,b\ldots$ using the \emph{fiducial} frame
$\ezero^a_i$ and co-frame $\ozero_a^i$ and indices are raised and
lowered with the fiducial metric $\qzero_{ab}$. Only purely
inhomogeneous smearing fields $\Lambda^b_{\inhom}$ contribute
because the first order perturbations are all purely inhomogeneous.
This equation involves $\aone_{ab}, \eone_{ab}$ linearly but also
contains the background variables $c,p$ which solve the zeroth order
constraint (\ref{homscalar1}). In this equation $(c,p)$ happen to
enter linearly. But in general the coefficients can be complicated
non-linear functions of the background fields. Thus, for example,
quadratic combinations appear in the first order vector constraint,
\be \sint N^a_{\inhom}\, [\ell\, p(\partial^b\aone_{ab} -
\partial_a \aone) - cp \mathring{\ep}_b{}^{mn}\aone_{mn} - c^2\ell
\mathring{\ep}_a{}^{mn}\eone_{mn} - \kappa\gamma \pphi \partial_a
\vpone]\, d\vzero\, = 0\, ,\ee
where $\a = \a_{ab}\qzero^{ab}$ is the trace of $\a_{ab}$. Note that
the first order constraints are linear `homogeneous' PDEs in the
sense that there are no (zeroth order) source terms on the right
hand side. The structure of the first order scalar constraint is the
same.

In the second order equations on the other hand are `inhomogeneous'
as PDEs since the zeroth and first order fields now act as
`sources'. These equations determine the `Coulombic' parts of the
second order fields in terms of the zeroth and first order ones. For
example, the second order Gauss constraint is given by
\be \label{gauss2} \sint \Lambda^b \, [ \ell \partial^a \etwo_{ab} +
c\, \mathring{e}_b{}^{mn}\, \etwo_{mn} - p\, \ell^{-1}
\mathring{\ep}_b{}^{mn}\atwo_{mn}]\, d\vzero = -\sint\, \Lambda^a\,
[\ell\mathring{\ep}_a{}^{bc} \aone_{db}\, \eone^{d}{}_c]\, d\vzero\,
. \ee
While this is again a linear PDE ---in fact the operator on the left
side is the same as that in (\ref{gauss1}), but now acts on the
second order fields, $(\atwo_a^i, \etwo^a_i)$--- there is now a
source term which is \emph{quadratic} in the first order fields
which are already known as solutions to (\ref{gauss1}).%
\footnote{Because the source terms on the right side of
(\ref{gauss2}) are quadratic in first order perturbations, it
follows that the integral on the right side does not necessarily
vanish if the smearing field $\Lambda^a$ is purely homogeneous. The
left side of (\ref{gauss2}) now implies that the second (and higher
order terms) in our $\ep$-expansion (\ref{expan}) \emph{cannot} be
assumed to be purely inhomogeneous.}
The structure of the second order vector and the scalar constraints
is the same. Finally, note that the $n$th order equations in this
hierarchy \emph{constrain only the $n$th order fields}; they do not
impose further conditions on lower order fields. This pattern
continues to all orders. In particular, to obtain the full set of
constraints on the zeroth and the first order fields, we need to
solve only\, $C|_{\ep =0} =0$ and $[dC/d\vp]|_{\ep =0} =0$. This
property will be important in what follows.

\emph{The key point about the hierarchy is that it greatly
simplifies the problem of solving the complicated, non-linear PDEs
$C(A,E; \Phi,\Pi) = 0$.} For $n=0$ we obtain a non-linear but
\emph{algebraic} equation. For $n>0$, each equation in the hierarchy
is a \emph{linear} PDE for the $n$th order fields, with the same
linear differential operator on the left side but order dependent
source terms which are already determined by solutions to the
\emph{lower order} equations in the hierarchy. The value of the
system lies in the hope that by truncating it to a low order, one
would obtain a good approximate solution to the full system with
small inhomogeneities. (For further discussion, see Appendix A.)
Although it is not easy to rigorously control the approximation,
this truncation scheme has proved to be a valuable and indispensable
tool in cosmological and black hole sectors of classical general
relativity. In practice, the domain of validity of the chosen
truncation is checked by \emph{self-consistency}: one only verifies
that, if the truncation is of order $n$, then the source terms in
equations governing $(n+1)$st order fields are negligibly small
compared to the fields that are kept.

What is the relation between this hierarchy of constraints and
dynamics? On the full phase space $(\ps, \Omega)$, dynamics is
generated by constraints. In particular, in the homogeneous sector,
the scalar constraint smeared with a homogeneous lapse can be
thought of as generating `pure time evolution'. What happens if we
truncate the theory at first order? Then, the truncated phase space
$\ps_{\tr}$ will be the normal bundle over $\ps_{\hom}$. The
Hamiltonian flow generated by $\mathbb{S}[N_{\hom}]$ on $\ps$ is
tangential to its homogeneous subspace $\ps_{\hom}$. It suffices to
consider this flow in an arbitrarily  small neighborhood of
$\ps_{\hom}$. Under this flow, tangent vectors $v \equiv (\aone_a^i,
\eone^a_i, \vpone, \pione)$ at any point on $\ps_{\hom}$ also have
an unambiguous evolution. Thus, given a specific $v_o$ at a point,
say $\gamma(t_o)$, of any dynamical trajectory $\gamma(t)$ on
$\ps_{\hom}$, the trajectory $\gamma(t)$ can be unambiguously lifted
to a trajectory in $\ps_{\tr}$, passing through $(\gamma(t_o),
v_o)$. We will see in section \ref{s3.4} that this lift has a simple
geometrical interpretation in the phase space.\\

\emph{Remark:} In the LQC literature, in place of the $\ep$
expansion (\ref{expan}) one often uses a decomposition of the type
$A_a^i = c\ell^{-1}\w_a^i + \delta A_a^i$, etc. Then the
`perturbations' $\delta A_a^i, \ldots$ include terms of all orders
$n\ge 1$ in our $\ep$ expansion. Therefore, in light of footnote 4,
it is not consistent to assume that these perturbations $\delta
A_a^i, \ldots$ are purely inhomogeneous. On the other hand, if one
allows them to have homogeneous parts, then the Poisson brackets
between the unperturbed and perturbed fields do not all vanish and
so the symplectic structure is more complicated. This complication
is often overlooked. In constraint equations, keeping terms linear
in these fields is interpreted as the first order truncation,
keeping terms quadratic as second order truncation, etc. In this
scheme, the second and higher order truncation would lead to
\emph{non-linear} PDEs on perturbations $\delta A_a^i, \ldots$;\,
the simplification that was achieved in the $\ep$-expansion
(\ref{expan}) would be lost. Truncations in the two schemes are
equivalent only to the linear order. These considerations apply also
to the Hamiltonian treatment of cosmological perturbations in terms
of metric variables, used outside the LQC literature as well.

\subsection{First order truncation}
\label{s3.3}

Since the temperature fluctuations in the cosmic microwave
background are only one part in $10^5$, much of the literature on
the early universe has focused on FLRW backgrounds with \emph{first
order} linear perturbations thereon. We will now analyze this
truncation in some detail. The relevant phase space is the normal
bundle $\ps_{\tr} = \ps_o \times \ps_1$ where $\ps_o = \ps_{\hom}$
is the 4-dimensional homogeneous subspace of the full phase space
$\ps$ and $\ps_1$ is spanned by the first order fields in the
expansion (\ref{expan}). It turns out that the description of the
FLRW \emph{quantum} geometries is easier in terms of variables $(\b,
\,\nu)$, rather than the original $(c,\, p)$:
\be \b = \f{c}{|p|^{1/2}}, \quad \nu =
\f{4|p|^{3/2}}{\kappa\gamma\hbar}\, {\rm sgn}\,\, p, \quad
{\hbox{\rm so that}}\quad \{\b,\, \nu \} = \f{2}{\hbar}\, . \ee
The geometrical meaning of these variables is as follows%
\footnote{Like $p$, the variable $\nu$ takes values in the entire
real line, positive values corresponding to triads $e^a_i$ with the
same orientation as the fiducial $\ezero^a_i$, and negative values
corresponding to triads with opposite orientation.}:
The physical volume of the universe is given by $a^3\ell^3 = 2\pi
\gamma\, |\nu|\, \lp^2$, where $a$ is the scale factor, and on any
solution $\b$ equals the Hubble parameter modulo a multiplicative
constant, $\b = \gamma (\dot{a}/a)$. Thus, we will now coordinatize
$\ps_o$ by $(\nu,\b; \phi,\pphi)$. For simplicity of notation,
\emph{from now on we will drop the suffix $(1)$ on the first order
perturbations.} Thus,
\be (\nu, \b,\, \phi, \pphi;\, \a_{ab}, \e_{ab},\,\vp,\pi) \in
\ps_{\tr}\, . \ee
The perturbations Poisson commute with the background fields and the
Poisson brackets among themselves are given by (\ref{pbs}) so that
$\Omega_{\tr} = \Omega_o + \Omega_1$. Thus, mathematically,
$(\ps_{\tr}, \Omega_{\tr}) = (\ps, \Omega)$. \emph{But the physical
interpretation of $\a_a^i, \e_a^i,\vp,\pi$ is different}: in
$\ps_{\tr}$ they represent only the first order perturbations, i.e,
just the coefficients of $\ep$ in the expansion (\ref{expan}). As a
result, the constraints and dynamical equations they satisfy are
very different from those of the full theory.

As noted in section \ref{s2.2}, to zeroth order we only have the
scalar constraint, smeared with a homogeneous lapse. In terms of
$\nu,b$ variables it becomes:
\be \mathbb{S}_o[N_{\hom}] = N_{\hom}\, [ -\f{3\hbar}{4\gamma}\,
\b^2\nu + \f{2\pphi^2}{\kappa\gamma\hbar\nu}]\, = \, 0. \ee
To first order in $\ep$, we obtain linear equations on first order
fields with functions of $\b,\nu, \pphi, \phi$ as coefficients. Let
us first focus on the Gauss constraint (\ref{gauss1}):
\be \label{gauss3} \sint \Lambda^b_{\inhom} \, [ \ell \partial^a
\eone_{ab} + c\, \mathring{\ep}_b{}^{mn}\, \eone_{mn} - p\,
\ell^{-1}\, \mathring{\ep}_b{}^{mn}\aone_{mn}] d\vzero\, =\,0\, .
\ee
It generates the following infinitesimal gauge transformations:
\be \a_{ab} \to \a_{ab} -\partial_a \Lambda_b + c\,\ell^{-1}
\mathring{\ep}_{abm} \Lambda^m, \quad \e_{ab} \to \e_{ab} + p\,
\ell^{-2} \mathring{\ep}_{abm} \Lambda^m \, . \ee
Thus, the symmetric part of $\e_{ab}$ is gauge invariant and
furthermore it has a simple interpretation. Since $E^a_i\, E^{bi} =
q q^{ab}$ in the full theory, where $q^{ab}$ is the physical
3-metric on $\man$ and $q$ its determinant, it follows that
\be \e_{(ab)} = - \f{\ell^2}{2p}\, (\q_{ab} - \q \,q^{(o)}_{ab}),
\quad {\rm where}\quad q_{ab} = q^{(o)}_{ab} + \ep \q_{ab} +
O(\ep^2)\, . \ee
(Note that $q^{(o)}_{ab}= (a^2)\, \qzero_{ab}$ is the zeroth order
\emph{physical} metric. It is purely homogeneous while $\q_{ab}$ is
purely inhomogeneous. $\q := \q_{ab} \qzero^{ab}$.) Next, recall
from (\ref{relation}) that $A_a^i(\ep) = \Gamma_a^i (\ep) - \gamma
K_a^i(\ep)$. Linearizing this equation one finds that the first
order Gauss constraint serves only to determine the skew symmetric
part of
\be \K_{ab} := \ozero_{bi}\, \f{d}{d\ep}\, K_a^i (\ep)|_{\ep=0} \ee
in terms of $\e_{ab}$ and the zeroth order fields. Thus imposition
of this constraint implies that $\K_{(ab)}$ is the free part of
$\K_{ab}$. Furthermore it is gauge invariant. Hence the reduced
phase space with respect to the first order Gauss constraint
(\ref{gauss1}) is coordinatized simply by the linearized metric
$\q_{ab}$ and the linearized extrinsic curvature $\K_{ab}$, both of
which are symmetric tensors. The three Gauss constraint have removed
three degrees of freedom from each of the two first order fields
$(\a_a^i, \e^a_i)$, taking us from the linearized connection
variables to the linearized ADM ones, $(\q_{ab}, \K_{ab})$. This is
precisely the structure one expects from the Gauss reduction of the
full theory \cite{aa-newvar}.

It now remains to impose the first order vector and scalar
constraints on the linearized pairs $(\q_{ab}, \K_{ab})$. For this
we can draw on the huge body of existing literature. Let us begin by
fixing the notation. The ADM variables are $(q_{ab}, p^{ab} =
\sqrt{q}\,(K_{ab} - K q_{ab}))$. To first order, they can be
expanded as
\be q_{ab}(\ep) = q^{(o)}_{ab} + \ep \q_{ab} + \ldots   \quad {\rm
and} \quad p^{ab}(\ep) = \sqrt{\qzero}\, (p^{(o)\,ab} + \ep \p^{ab}
+ \ldots ) \ee
where the zeroth order fields are given by $q^{(o)}_{ab} = a^2
\qzero_{ab}$ and $p^{(o)\,ab} = -(ab/\kappa\gamma)\, \qzero^{ab}$.
As in the cosmology literature, we can expand out the first order
fields using Fourier transforms:
\ba \q_{ab}(\x) = \f{1}{\Vzero}\, \sum_{\vk\in {\mathcal{L}}}
\tilde{\q}_{ab}(\vk)\, e^{i\vk\cdot\x} &\quad\quad& \vp(\x)=
\f{1}{\Vzero}\, \sum_{\vk\in {\mathcal{L}}}{\vp}_{\vk}\,
e^{i\vk\cdot\x}\nonumber\\
\p_{ab}(\x) = \f{1}{\Vzero}\, \sum_{\vk\in {\mathcal{L}}}
\tilde{\p}_{ab}(\vk)\, e^{i\vk\cdot\x} &\quad\quad& \pi(\x)=
\f{1}{\Vzero}\, \sum_{\vk\in {\mathcal{L}}}\,\pi_{\vk}\,
e^{i\vk\cdot\x}\,  \ea
where $\mathcal{L}$ is the lattice defined by $\vk \in ((2\pi/\ell)
\mathbb{Z})^3,\, \vk \not= \vec{0}$,\,\, $\mathbb{Z}$ being the set
of integers. (The zero $\vk$ is excluded because, by construction,
the fields are all purely inhomogeneous.)
%
%\ba \q_{ab}(\x) = \textstyle{\f{1}{(2\pi)^3}}\,\sint d^3k\,
%\tilde{\q}_{ab}(\vk)\, e^{i\vk\cdot\x} &\quad\quad& \vp(\x)=
%\textstyle{\f{1}{(2\pi)^3}}\,\sint d^3k\, {\vp}_{\vk}\,
%e^{i\vk\cdot\x}\nonumber\\
%\p_{ab}(\x) = \textstyle{\f{1}{(2\pi)^3}}\,\sint d^3k\,
%\tilde{\p}_{ab}(\vk)\, e^{i\vk\cdot\x} &\quad\quad& \pi(\x)=
%\textstyle{\f{1}{(2\pi)^3}}\,\sint d^3k\, \,\pi_{\vk}\,
%e^{i\vk\cdot\x}\, \ea
%
Since all four fields in the $\x$ space are real, their Fourier
transforms satisfy the relations $\tilde{\q}_{ab}(\vk) =
\tilde{\q}_{ab}^\star(-\vk)$, etc. It is convenient to expand out
the metric perturbations into scalar, vector and tensor modes
\cite{bardeen}:
\ba \tilde{\q}_{ab}(\vk) =&& S^{(1)}_{\vk}\,\qzero_{ab} +
S^{(2)}_{\vk} (\hat{k}_a\hat{k}_b - \f{1}{3} \qzero_{ab}) + \sqrt{2}
V^{(1)}_{\vk}\, \hat{k}_{(a}\hat{x}_{b)}\nonumber\\
&+& \sqrt{2} V^{(2)}_{\vk}\, \hat{k}_{(a}\hat{y}_{b)} +
\f{1}{\sqrt{2}}\, T^{(1)}_{\vk} (\hat{x}_a\hat{x}_b -
\hat{y}_a\hat{y}_b) + \sqrt{2} T^{(2)}_{\vk}\,
(\hat{x}_{(a}\hat{y}_{b)}) \ea
and, \ba \tilde{\p}_{ab}(\vk) =&& \f{1}{3}\,
\p^{(S_1)}_{\vk}\,\qzero_{ab} + \f{1}{2}\, \p^{(S_2)}_{\vk}
(3\hat{k}_a\hat{k}_b - \qzero_{ab}) + \sqrt{2}
\p^{(V_1)}_{\vk}\, \hat{k}_{(a}\hat{x}_{b)}\nonumber\\
&+& \sqrt{2} \p^{(V_2)}_{\vk}\, \hat{k}_{(a}\hat{y}_{b)} +
\f{1}{\sqrt{2}}\, \p^{(T_1)}_{\vk} (\hat{x}_a\hat{x}_b -
\hat{y}_a\hat{y}_b) + \sqrt{2} \p^{(T_2)}_{\vk}\,
(\hat{x}_{(a}\hat{y}_{b)})\,. \ea
(Here $\hat{k}$ is a unit vector in the $\vk$ direction and
$\hat{k}, \hat{x},\hat{y}$ constitutes a field of orthonormal triads
with respect to ${\qzero}_{ab}$ in the momentum space. Throughout,
indices are lowered, raised and contracted using $\qzero_{ab}$.)
Then, the canonically conjugate pairs are $(S^{(1)}_{\vk},
\p^{S_1}_{-\vk}),\,\, (S^{(2)}_{\vk}, \p^{S_2}_{-\vk}),\,\, \ldots
\,\, (\vp_{\vk}, \pi_{-\vk})$, with Poisson brackets
$\{S^{(1)}_{\vk},\, \p^{S_1}_{-\vk^\prime}\}  = \ell^3
\delta_{\vk,\,\vk^\prime}, \ldots\, $. These fields are subject to
three vector constraints and a scalar constraint, each of which is
linear in these fields but also contains background fields. As is
well known, one can pass to the reduced, truncated phase space
$\tilde{\Gamma}^{(1)}_{\tr}$ by solving them and finding gauge
invariant variables \cite{langlois,puchta}:
\be \tilde\Gamma^{(1)}_{\tr}\, = \, \{ (\Q_{\vk}, T^{(1)}_{\vk},
T^{(2)}_{\vk};\, \p^{(\Q)}_{\vk}, \p^{(T_1)}_{\vk},
\p^{(T_2)}_{\vk})\}\, .\ee
Here $T^{(1)}_{\vk}$ and $ T^{(2)}_{\vk}$ are the two tensor modes
(which are automatically gauge invariant) and $\Q_{\vk}$, the gauge
invariant Mukhanov-Sasaki variable, is given by
\be \Q_{\vk} = \vp_{\vk} - \f{\pphi \gamma}
{2 a^5 \ell^3 \b} \, \left(S^{(1)}_{\vk} -
\f{1}{3}\, S^{(2)}_{\vk}\right)\, . \ee
Its conjugate momentum is given by
\be \p^{\Q}_{\vk} = \pi_{\vk} + \frac{\kappa \gamma \pphi^2}{2
\ell^6 a^3 b} \, \vp_{\vk} -\frac{\kappa \gamma^2 \pphi^3}{4 \ell^9
a^8 b^2} \, S^{(1)}_{\vk}-\left(\frac{\pphi}{2
a^2\ell^3}-\frac{\kappa \gamma^2 \pphi^3}{12 \ell^9 a^8 b^2}\right)
S^{(2)}_{\vk} \, . \ee
The initial configuration variable $\q_{ab}$ had six degrees of
freedom and the scalar field $\vp$, one. Each of the four linearized
constraints reduces the configuration degrees of freedom by one,
leaving us with 3 degrees of freedom $(\Q, T^{(1)}, T^{(2)})$ in the
reduced configuration space.

Let us summarize. Note first that in the passage to the reduced
phase space $\t{\ps}_{\tr}$, we used only the kinematical structure
on $\ps_{\tr}$; \emph{dynamical equations were not used.} Thus, the
procedure is completely analogous to that followed in other LQC
models to extract the physical degrees of freedom (see \ref{s2.2}).
$\tilde\Gamma_{\tr}$ has the form
\be \tilde\Gamma_{\tr} = \Gamma_o \times \tilde{\Gamma}^{(1)}, \quad
{\rm with} \quad (\nu, \b; \phi, \pphi)\in \Gamma_o, \,\,\,
(\Q_{\vk}, T^{(1)}_{\vk}, T^{(2)}_{\vk};\, \p^{\Q}_{\vk},
\p^{T_1}_{\vk}, \p^{T_2}_{\vk}) \in \tilde{\Gamma}^{(1)}\ee
These variables are subject to only the zeroth order scalar
constraint:
\be \label{scalaro} \mathbb{S}_o[N_{\hom}] = N_{\hom}\, [
-\f{3\hbar}{4\gamma}\, \b^2\nu +
\frac{2\pphi^2}{\kappa\gamma\hbar\nu}]\, =\, 0\, .\ee
We have already taken care of the first order constraints. As noted
above, the second and higher order constraints do not restrict the
first order variables.\medskip

\emph{Remarks:}

1. In the CMB, we can only observe modes up to a (finite) maximum
wave length $\lambda_o$ which equals the radius of the observable
universe at the surface of last scattering. Therefore, it is
physically appropriate to absorb modes with wave-lengths $\lambda
\gtrsim 10\lambda_o$ in the homogeneous background. This amounts to
putting a \emph{physical} infrared cut-off on perturbative modes,
making the arbitrariness in the choice of the radius of the 3-torus
$\mathbb{T}^3$ irrelevant. In phenomenological applications, these
considerations have to be folded into the calculation of the
renormalized energy density and checking self-consistency.

2. We use the Mukhanov-Sasaki variable $\Q_{\vk}$ ---rather than the
curvature perturbation $\R_{\vk}$--- to coordinatize the reduced
phase space because it is better suited for our discussion of
inflation in \cite{aan3}. There, we find that $\R_{\vk}$ is not
well-defined along effective trajectories because it carries $\pphi$
in the denominator and $\pphi$ vanishes at `turning points' of the
inflaton. $\Q_{\vk}$ on the other hand is well defined all along
these trajectories. In absence of a potential $V(\phi)$, as in this
paper, they are related just by a constant: $\R_{\vk} =
\sqrt{\kappa/6}\, \Q_{\vk}$.

\subsection{Truncated Dynamics}
\label{s3.4}

As mentioned in section \ref{s3.1}, to obtain dynamical trajectories
on $\tilde\Gamma_{\tr}$ one has to proceed in two steps: First
obtain a dynamical trajectory on $\Gamma_o$ and then lift it to
$\tilde{\Gamma}_{\tr}$. On $\Gamma_o$, dynamics is generated by the
scalar constraint $\mathbb{S}[N_{\hom}]$, i.e., the Hamiltonian
vector field is just the restriction to $\Gamma_o$ of the full
Hamiltonian vector field%
\footnote{Here the Greek indices refer to the (infinite dimensional)
tangent space to the full or truncated phase space. They are to be
regarded as abstract indices a la Penrose. This index notation can
be avoided; it is used only as a pedagogical aid.}
$X^\alpha = \Omega^{\alpha\beta}\partial_\beta \mathbb{S}[N_{\hom}]$
on $\Gamma$:\,\, $X^\alpha|_{\Gamma_o} =
\Omega_o^{\alpha\beta}\,\,\partial_\beta \mathbb{S}_o[N_{\hom}]$.
Full dynamics on $\Gamma$ unambiguously induces a flow on
$\tilde\Gamma_{\tr}$. The general procedure for non-linear systems,
summarized in Appendix A, now implies that the time evolution on the
truncated phase space $\Gamma_{\tr}$ is given by the dynamical
vector field $X^\alpha_{\dyn}$,
\be \label{dynX} X^\alpha_{\dyn}=
\Omega_o^{\alpha\beta}\,\partial_\beta \mathbb{S}_o[N_{\hom}]
+\Omega_{1}^{\alpha\beta}\partial_\beta
\mathbb{S}^\prime_{2}[N_{\hom}]\, ,\ee
where $\mathbb{S}^\prime_2$ is obtained from the coefficient
$\mathbb{S}_2[N_{\hom}]$ of $\ep^2$ in the expansion of the full
scalar constraint $\mathbb{S}[N_{\hom}]$ on $\Gamma$, by keeping
just those terms which are quadratic in the first order quantities
$(\a^{(1)}_a{}^i, \e^{(1)a}{}_{i}, \vp^{(1)}, \pi^{1})$ and
discarding terms linear in the second order quantities
$(\a^{(2)}_a{}^i, \e^{(2)a}{}_{i}, \vp^{(2)}, \pi^{(2)})$. It is
important to note that it is only $\mathbb{S}_2[N_{\hom}]$ that is
constrained to vanish;\, \emph{there is no constraint on}
$\mathbb{S}^\prime_2[N_{\hom}]$. Finally, because
$\mathbb{S}^\prime_2$ depends on the background fields,
$X^\alpha_{\dyn}$ is not generated by $\mathbb{S}_o +
\mathbb{S}^\prime_2$; i.e., $X^\alpha_{\dyn} \not=
\Omega^{\alpha\beta}\, \partial_\beta(\mathbb{S}_o
+\mathbb{S}^\prime_2)$. But one might wonder if one can find another
`effective Hamiltonian' that generates this dynamical flow. The
answer is in the negative: One can check that $X^\alpha_{\dyn}$
\emph{does not Lie drag the total symplectic structure $\Omega =
\Omega_o+\Omega_1$ on the truncated phase
space}, whence it is impossible to find an effective Hamiltonian.\\

\emph{Remark:} We have spelled out these results because there has
been some confusion in the recent LQC literature. \emph{Quantum}
dynamics is often assumed to be captured by a quantum constraint
$\hat{\mathbb{\underbar{C}}}_H \Psi =0$. Our detailed discussion of
truncation and dynamics of the truncated system implies that this
assumption cannot be justified. First, as Eq. (\ref{dynX}) shows,
the dynamical vector field $X^\alpha_{\dyn}$ on $\Gamma_{\tr}$ is
not generated by a constraint (or indeed by \emph{any} Hamiltonian).
Therefore there is no reason to expect that the correct quantum
dynamics could be recovered by imposing any quantum constraint.
Second, even the part $\Omega_{1}^{\alpha\beta}\partial_\beta
\mathbb{S}^\prime_2[N_{\hom}]$ of $X^\alpha_{\dyn}$ describing the
evolution of perturbations is \emph{not} generated by the second
order constraint $\mathbb{S}_2[N_{\hom}]$ but by
$\mathbb{S}^\prime_2[N_{\hom}]$, which is unconstrained.\\

Finally, we are interested only in the dynamics of gauge invariant
variables, i.e., only in dynamics on $\tilde{\Gamma}_{\tr}$. As one
would expect, the functions $\mathbb{S}_o$ and $\mathbb{S}^\prime_2$
that determine $X^\alpha$ are gauge invariant and therefore projects
down to $\tilde{\Gamma}_{\tr}$. In particular, $\mathbb{S}^\prime_2
= \mathbb{S}^\prime_2{}^{(\Q)} + \mathbb{S}^\prime_2{}^{(T_1)} +
\mathbb{S}^\prime_2{}^{(T_2)}$ and the form of the last two terms is
exactly the same. \emph{ Therefore, from now on, we will denote the
two tensor modes collectively by} $\T$. Then \cite{langlois},
\ba \label{pert-ham} \mathbb{S}^\prime_2{}^{(\Q)}\,[N_{\hom}] &=&
\f{N_{\hom}}{2\Vzero}\,\, \sum_{\vk} \f{1}{a^3}\,
|\p^{(\Q)}_{\vk}|^2 + {a} k^2\, |\Q_{\vk}|^2 \nonumber\\
\mathbb{S}^\prime_2{}^{(\T)}\,[N_{\hom}] &=&
\f{N_{\hom}}{2\Vzero}\,\, \sum_{\vk} \f{4\kappa}{a^3}\,
|\p^{(\T)}_{\vk}|^2 + \f{ak^2}{4\kappa}\,|\T_{\vk}|^2\, . \ea
%
%where $z = \sqrt{6/\kappa}\, a$. In presence of a potential for the
%scalar field, the expressions of the Hamiltonian remain the same but
%the definition of the variable $z$ used in the Hamiltonian of the
%scalar perturbation is modified (See Appendix B). This is the
%motivation behind introducing $z$. Substituting for $z$, we find that if
%
Note that if we were to replace $2\sqrt{\kappa}\Q_k$ by $\T_k$ (and
the corresponding momenta by a reciprocal factor to maintain the
Poisson brackets), $\mathbb{S}^\prime_2{}^{(\Q)}\,[N_{\hom}]$
reduces to $\mathbb{S}^\prime_2{}^{(\T)}\,[N_{\hom}]$.
\emph{Therefore, it suffices to focus just on (one of the two tensor
modes) $\T_{\vk}$ and its conjugate momentum which we will denote
simply by $\p_{\vk}$}.

The resulting dynamical evolution can  be best understood in
geometric terms as follows. Note first that $\t\Gamma_{\tr}$ is
naturally a bundle over the homogeneous phase space $\Gamma_o$. Fix
an integral curve of $X^\alpha$ on the `base space' $\Gamma_o$, i.e.
a homogeneous solution. Fix a point $\gamma_o \equiv
(\nu^o,\b^o;\,\phi^o, \pphi^o)$ on this trajectory and a tangent
vector $\T^o_{\vk}, \p^o_{\vk}$ at that point. To describe the
evolution of this perturbation on the chosen background trajectory,
we need to lift the trajectory from $\Gamma_o$ to
$\tilde{\Gamma}_{\tr}$. This can be done simply by considering the
integral curve of $X^\alpha_{\dyn}$ passing through the point
$(\nu^o,\b^o;\,\phi^o,\pphi^o;\, \T^o_{\vk}, \p^o_{\vk})$ of
$\tilde{\Gamma}_{\tr}$.\\

Next, let us list the commonly used lapse functions $N_{\hom}$ and
the corresponding time variables:
\begin{itemize}
\item $N_{\hom}= 1$ corresponds to proper time $t$ so that the
    physical space-time metric has the form $ds^2 = - dt^2 + a^2
    d\x^2$.
\item $N_{\hom}= a$ corresponds to conformal time $\eta$ so that
    the physical space-time metric has the form $ds^2 = a^2(-
    d\eta^2 + d\x^2)$. This is the most common choice in the
    cosmology literature.
\item $N_{\hom}= a^3$ corresponds to harmonic time $\tau$ which
    satisfies the wave equation, $\Box \tau =0$. The physical
    space-time metric now assumes the form $ds^2 = - a^6 d\tau^2
    + a^2 d\x^2$. The zeroth order scalar constraint
    $\mathbb{S}_o [N_{\hom}]$ takes the simplest form with this
    choice. Therefore, harmonic time is commonly used in LQC not
    only for the $k\!=\!0,\, \Lambda\!=\!0$ FLRW cosmology but
    also for models that admit spatial curvature, a non-zero
    cosmological constant and anisotropies
    \cite{acs,awe2,awe3,we,asrev}.
\item $N= (\Vzero/\pphi) a^3$ corresponds to choosing the scalar
    field $\phi$ itself as time. It turns out that irrespective
    of the initial choice of lapse in the classical theory, the
    \emph{quantum} scalar constraint has a form that naturally
    leads one to use $\phi$ as a relational or internal time
    variable \cite{aps3,acs}. In the case now under
    consideration, where the scalar field $\phi$ also satisfies
    the wave equation $\Box \phi =0$, the internal time defined
    by $\phi$ and the harmonic time $\tau$ are related just by a
    constant in any classical solution: $\phi = (\pphi/\Vzero)
    \tau$. However, since the constant varies from one solution
    to another, in the quantum theory, conceptually, the two
    choices are quite different. In LQC it is simplest to begin
    with the harmonic time in the classical theory and
    reinterpret the quantum scalar constraint as providing time
    evolution in the relational time $\phi$.
\end{itemize}

Since one generally uses the conformal time $\eta$ in the cosmology
literature, we will conclude by writing down the equation of motion
that follows from (\ref{pert-ham}) with the choice $N_{\hom} = a$:
\be \label{eomT} \T^{\prime\prime}_{\vk}  + 2\, \f{a^\prime}{a}\,
\T^\prime_{\vk} + k^2 \T_{\vk} = 0 \ee
where a `prime' denotes derivative with respect to the conformal
time $\eta$. One often rescales $\T_{\vk}$ to obtain a field
$\chi_{\vk}$ with physical dimensions of a scalar field,
\be \chi_{\vk} := \f{a}{2\sqrt{\kappa}}\, \T_{\vk}\,  ,  \ee
for which, furthermore, the equation of motion resembles that of a
harmonic oscillator (with time dependent frequency):
\be \label{eom} \chi_{\vk}^{\prime\prime} +
(-\frac{a^{\prime\prime}}{a} + k^2)\, \chi_{\vk}\,\, \equiv \,\,
\chi_{\vk}^{\prime\prime} + a^2(-\f{R}{6} + \f{k^2}{a^2})\,
\chi_{\vk} =0\, ,\ee
where $R$ denotes the scalar curvature of the background homogeneous
space-time.%
\footnote{For the scalar perturbation, $\psi_{\vk} = a\Q_{\vk}$ has
the same properties as $\chi_{\vk}$:  $\psi_{\vk}$ also has the
physical dimensions of a scalar field and satisfies the same
equation of motion as (\ref{eom}).}
This form of the perturbation equations has also been exploited in
\cite{iberian,madrid2,madrid3} in singling out a physically
motivated quantization.

\emph{Remark:} In our presentation we began with the connection
variables employed in the LQG literature because conceptually they
are essential for the new quantum kinematics used in LQG, and the
subsequent treatment of the singularity-free quantum dynamics of our
background FLRW space-times. But we passed to the ADM variables for
first order perturbations by solving the Gauss constraint because
much of the cosmological perturbation analysis is carried out in
terms of these variables. For a treatment of cosmological
perturbations in connection variables, see in particular
\cite{puchta,dt,pert_scalar,ghtw1,ghtw2,joao2}.

\section{Quantum theory: Main Steps}
\label{s4}

We will now use the phase space $\tilde{\Gamma}_{\tr}$ as the point
of departure for quantization. Since the truncated second order
constraints for the scalar and two tensor modes are identical
(except for unimportant numerical factors), as in section \ref{s3.4}
we will focus just on one tensor mode $\T_{\vk}$. Thus, from now on,
$\tilde{\Gamma}_{\tr}$ will be taken to be spanned by three
canonically conjugate pairs  $(\nu,b;\, \phi,\pphi;\,\, \T_{\vk},
\p_{\vk})$, the first two representing the background, and the third
representing the first order perturbation. They are subject to the
single constraint $\mathbb{S}_o[N_{\hom}] = 0$ (see Eq
(\ref{scalaro})). Dynamics is governed by the vector field
$X^\alpha_{\dyn} = \Omega_o^{\alpha\beta}\partial_\beta
\mathbb{S}_o+
\Omega_1^{\alpha\beta}\partial_\beta\mathbb{S}^\prime_2$ where
$\mathbb{S}^\prime_2$ is given by (\ref{pert-ham}).

In this section we present the general program without entering into
details of how the states and operators related to first order
perturbations are defined. In particular, the expression of the
Hamiltonian operator dictating the dynamics of first order
perturbations is formal. The precise definitions of these states,
operators and the necessary regularization procedure are provided in
the next two sections

\subsection{Background quantum geometry} \label{s4.1}

Let us begin by recalling the quantum theory of the homogeneous
sector (for details, see e.g. \cite{aps3,acs,asrev}). As mentioned
in section \ref{s2}, LQC provides a kinematical Hilbert space
$\Hkin$ which, as in full LQG \cite{lost,cf} is uniquely determined
by the physical requirement independence w.r.t. background (and
fiducial) structures \cite{ac}. In the configuration representation,
kinematical quantum states are given by wave functions
$\Psi(\nu,\phi)$. As noted in section \ref{s3.4}, dynamics is
simplest if one uses harmonic time by setting $N=a^3$. Then, the
scalar constraint becomes a well-defined operator on $\Hkin$.
Physical states $\Psi_o(\nu,\phi)$ are annihilated by this
constraint, i.e., they satisfy
\be \label{qham1}\hat{\mathbb{S}}_o \Psi_o(\nu,\phi) \equiv
-\f{\hbar^2}{2\ell^3}\,\,(\partial_\phi^2 + \Theta) \Psi_o(\nu,
\phi) = 0\, , \ee
where the action of $\Theta$ is given by
\be \label{Theta} \Theta \Psi_o(\nu,\phi) = \f{3\pi G}{\lambda^2}\,
\nu \big[(\nu+2\lambda) \Psi_o(\nu+4\lambda, \phi) - 2\nu
\Psi_o(\nu,\phi) + (\nu-2\lambda)\Psi_o(\nu-4\lambda, \phi) \big]\, .
\ee
Thus, $\Theta$ is a second order difference operator that acts only
on the argument $\nu$ of $\Psi_o$, with step size $4\lambda$ where
$\lambda^2 = 4\sqrt{3}\pi\gamma\lp^2$ is the `area gap' of LQG.
$\Theta$ is self-adjoint and positive definite on $\Hkin$
\cite{kl1}. As is common in the Dirac quantization procedure, none
of the solutions to (\ref{qham1}) are normalizable in the kinematic
inner product. But there is a standard `group averaging' method to
endow the space of solutions with a physical inner product
\cite{dm,almmt,abc}. The resulting physical states $\Psi_o \in \Hpo$
turn out to be solutions to
\be \label{qham2} -i\hbar\,\partial_\phi \Psi(\nu,\phi) = \hat{H}_o
\Psi_o(\nu,\phi)\quad\quad {\rm where} \quad \hat{H}_o =
\hbar\,\sqrt{\Theta}\, , \ee
which are symmetric under $\nu \rightarrow -\nu$ and have finite
norm
\be ||\Psi_o||^2 = \f{\lambda}{\pi}\,\,\sum_{\nu\in 4N\lambda;\,\,
N\in \mathbb{Z}} |\nu|^{-1}\,\,|\Psi_o(\nu, \phi_o)|^2\, ,
\label{norm}\ee
where $\phi_o$ is any fixed instant of the internal time $\phi$.
(The scalar product defined by (\ref{norm}) is insensitive to the
choice of $\phi_o$.) Heuristically (\ref{qham2}) can be thought of
as the positive frequency square-root of (\ref{qham1}) in the
internal time $\phi$.

The most noteworthy feature of this outcome is that the
\emph{quantum} Hamiltonian constraint is naturally de-parameterized:
its form suggests that the scalar field $\phi$ can be interpreted as
a `relational or internal time' with respect to which physical
states $\Psi_o$ evolve. Thus, as one would hope, the imposition of
the quantum constraint a la Dirac has naturally led us to dynamics.
Interestingly, while our use of a lapse function corresponding to
the harmonic time $\tau$ simplified the form of $\hat{H}_o$, it was
not essential to arrive at this interpretation of $\phi$. Indeed
this interpretation was initially arrived at with lapse set to $1$
corresponding to proper time \cite{aps3} (but a more complicated
form of $\hat{H}_o$.)

Thus, as is usual, the Dirac quantization procedure has naturally
led to a Schr\"odinger picture in which the scalar field $\phi$ is
simply a time parameter and the sole dynamical variable is $\nu$
that determines the volume of the universe via
\be \label{V2} \h{V}\Psi_o(\nu) = 2\pi \gamma \lp^2 |\nu|
\Psi_o(\nu)\, .\ee
In the Heisenberg picture, the volume evolves in time from the
bounce via
\be \hat{V}(\phi) = e^{(i/\hbar)\h{H}_o (\phi-\phi_{\rm{B}})}\,\,
(2\pi \gamma\lp^2 |\nu|)\,\, e^{-(i/\hbar)\h{H}_o (\phi-\phi_{
\B})}\, .\ee

Every element $\Psi_o$ of $\Hpo$ represents a 4-dimensional quantum
geometry. However, for our purposes, only a subset of these states
is relevant.  Choose a classical, expanding FLRW space-time in which
$\pphi \gg \hbar$ (in the classical units $G$=$c$=1) and a
homogeneous slice at a late time $\phi = \phi_o$, when the matter
density and curvature are negligibly small compared to the Planck
scale. This defines a point $\gamma_o$ in $\Gamma_o$. Then, in the
Schr\"odinger representation, one can introduce `coherent states'
$\Psi_o (\nu, \phi_o)$ in $\Hpo$ which are sharply peaked at
$\gamma_o$ \cite{aps2}. \emph{By a quantum background geometry, we
will mean a physical state $\Psi_o(\nu,\phi)$ obtained by evolving
these initial states using (\ref{qham2}).} There is a large class of
such states and our considerations will apply to all of them. One
can show that these states remain sharply peaked on the classical
trajectory passing through $\gamma_o$ for all $\phi
> \phi_o$. In the backward time-evolution, they do so only till the
density reaches a few hundredths of the Planck density. Even in the
deep Planck regime the wave function remains sharply peaked but now
the peak follows an effective trajectory which undergoes a quantum
bounce. At the bounce point the matter density attains a maximum,
$\rho_{\rm max} \approx 0.41 \rho_{\rm Pl}$. While there is good
agreement with general relativity once the matter density falls
below a few hundredths of the Planck density, Einstein's equations
break down completely in the Planck regime. But the quantum state
(and even the effective trajectory) remains well-defined throughout
the entire evolution, including the Planck scale neighborhood of the
bounce \cite{aps3,acs,asrev}.

Finally, in the Heisenberg picture, we can define the space-time
metric operator. Recall first that in the classical theory the lapse
function corresponding to the scalar field time is given by $N =
a^3(\phi)\,\ell^3\, \pphi^{-1}$. Since $\hat{p}_{(\phi)} \Psi_o =
\hat{H}_o\Psi_o$ for any $\Psi_o \in \Hpo$, in the Heisenberg
representation, the metric operator is given by
\be \label{gop} \hat{g}_{ab}dx^a dx^b \, \equiv\,  d\hat{s}^2 \,= \,
\hat{H}_o^{-1}\, {\ell^6\,\hat{a}^6(\phi)}\,\hat{H}_o^{-1}\, d\phi^2
\,+\,\hat{a}^2(\phi)\,\,d\x^2 \ee
where we have used a symmetric factor ordering in the first term and
defined the (positive definite, self-adjoint) scale factor operator
in the Heisenberg picture via:
\be \ell \hat{a}(\phi) = [\hat{V}(\phi)]^{\f{1}{3}}\, .  \ee
In the Heisenberg picture, the geometry is quantum because the
metric coefficients are now quantum operators on $\Hpo$.

\subsection{Perturbations on the quantum geometry $\Psi_o$}
\label{s4.2}

Because we were able to reinterpret the quantum constraint equation
$\h{\mathbb{S}}_o\Psi_o =0$ as providing an evolution of physical
states in the internal time variable $\phi$, we were naturally led
to work in the Schr\"odinger picture for the homogeneous background
geometry. Furthermore, because $\Psi_o$ represents a quantum state
of the background geometry and $\psi$ of perturbations, it is
natural to assume that the total state has a simple tensor product
structure
\be \Psi(\nu,\T_{\vk}, \phi_{0}) = \Psi_o(\nu,\phi_{0}) \otimes
\psi(\T_{\vk}, \phi_{0}) \ee
at some initial time $\phi_0$. Then, because the back-reaction is
neglected, the evolution of $\Psi_o$ is dictated just by
(\ref{qham2}); it is insensitive to the form of $\psi$. (This is
entirely analogous the situation in the classical theory.) Therefore
the tensor product structure is preserved under evolution. As in the
classical theory, our task is to evolve $\psi$ on the specified
background quantum geometry $\Psi_o$, i.e., to lift the given
homogeneous `quantum trajectory' $\Psi_o(\nu,\phi)$ to a
`trajectory' $\Psi_o(\nu,\phi) \otimes \psi(\T_{\vk},\phi)$ of the
truncated quantum theory, where the background state is the given
one. To carry out this task we need to first complete two
preliminary steps.

On the classical phase space $\tilde{\Gamma}_{\tr} = \Gamma_o\times
\tilde\Gamma^{(1)}$, the part $\Omega^{\alpha\beta}_1\partial_\beta
\mathbb{S}^\prime_2$ of the dynamical vector field $X^\alpha_{\dyn}$
dictates how perturbations propagate on a homogeneous background
solution $\nu(\phi)$. Now, $\mathbb{S}_2^\prime$ depends not only on
the perturbations $(\T_{\vk}, \p_{\vk})$ but also on the \emph{time
dependent scale factor of the background solution}. Therefore, to
construct the operator $\h{\mathbb{S}}_2^\prime$, it is simplest to
work in the `interaction picture' where the background scale factor
operators evolve in the relational time $\phi$ and the background
state $\Psi_o$ is frozen at a time, which we will take to be the
bounce time $\phi= \phi_{\B}$. The first preliminary step is to
carry out this passage to the interaction picture via
\be \Psi_{\inter} (\nu,\T_{\vk},\phi) = e^{-(i/\hbar)\, \hat{H}_o
(\phi-\phi_{\B})}\,\, \big(\Psi_o(\nu,\phi) \otimes \psi(\T_{\vk},
\phi)\big)\, . \ee
A second step is needed because now the evolution is with respect to
the relational time $\phi$. Therefore, we have to choose a specific
lapse function, $N_{\hom} = a^3\ell^3/\pphi$, in the expression
(\ref{pert-ham}) of $\mathbb{S}_2^\prime$ and then use an
appropriate factor ordering to convert it to an operator. In this
step, we will use the same factor ordering as in the expression
(\ref{gop}) of the quantum metric operator and make a simplification
using the evolution equation $\hat{p}_{(\phi)} \Psi_o \equiv -i\hbar
\partial_\phi \Psi_0 = \h{H}_0\Psi_0$.

These two steps, and the form (\ref{pert-ham}) of
$\mathbb{S}_2^\prime$ lead us to the following evolution equation
for the total state in the interaction picture:
\ba i\hbar\partial_\phi \Psi_{\inter} (\nu,\T_{\vk},\phi) &=&
\Psi_o(\nu,\phi_{\B}) \otimes i\hbar\partial_\phi
\psi(\T_{\vk},\phi)\nonumber\\
&=& \f{1}{2}\, \sum_{\vk} 4\kappa
[\hat{H}_o^{-1}\,\Psi_o(\nu,\phi_{\B})]\otimes
[|\hat{\p}_{\vk}|^2\psi(\T_{\vk},\phi)]\nonumber\\
&+& \f{k^2}{4\kappa} [(\hat{H}_o^{-\f{1}{2}}\, \hat{a}^4(\phi)\,
\hat{H}_o^{-\f{1}{2}}) \Psi_o(\nu,\phi_{\B})] \otimes
[|\h\T_{\vk}|^2 \psi(\T_{\vk}, \phi)]\, . \label{qevo1}\ea
%
%
%\ba i\hbar\partial_\phi \Psi_{\inter} (\nu,\T_{\vk},\phi) &=&
%\Psi_o(\nu,\phi_{\B}) \otimes i\hbar\partial_\phi
%\psi(\T_{\vk}\phi)\nonumber\\
%&=& \f{1}{2}\, \f{\ell^3}{(2\pi)^3}\, \int\! d^3k\,\,\Big[\, 4\kappa
%[\hat{H}_o^{-1}\,\Psi_o(\nu,\phi_{\B})]\otimes
%[|\hat{\p}_{\vk}|^2\psi(\T_{\vk},\phi)]\nonumber\\
%&+& \f{k^2}{4\kappa} [(\hat{H}_o^{-\f{1}{2}}\, \hat{a}^4(\phi)\,
%\hat{H}_o^{-\f{1}{2}}) \Psi_o(\nu,\phi)] \otimes [|\T_{\vk}|^2
%\psi(\T_{\vk}, \phi)]\, \Big] \label{qevo1}\ea
%
Let us take the scalar product of this equation with $\Psi_o$ (which
we assume to be normalized). the result is the required evolution
equation for the quantum state $\psi(\T_{\vk}, \phi)$ of the
perturbation propagating on the quantum background geometry
$\Psi_o$:
\ba i\hbar\partial_\phi \psi(\T_{\vk}, \phi) &=& \hat{H}_1
\psi(\T_{\vk}, \phi) := \f{1}{2}\, \sum_{\vk} 4\kappa \langle
\hat{H}_o^{-1}\rangle\,
|\hat{\p}_{\vk}|^2 \psi(\T_{\vk},\phi)\nonumber\\
&+& \f{k^2}{4\kappa}\, \langle \hat{H}_o^{-\f{1}{2}}\,
\hat{a}^4(\phi)\, \hat{H}_o^{-\f{1}{2}}\rangle\, |\h\T_{\vk}|^2
\psi(\T_{\vk},\phi)\, , \label{qevo2}\ea
%
%\ba i\hbar\partial_\phi \psi(\T_{\vk}, \phi) &=& \hat{H}_1
%\psi(\T_{\vk}, \phi) := \f{1}{2}\, \f{\ell^3}{(2\pi)^3}\, \int\!
%d^3k\,\,\Big[\, 4\kappa \langle \hat{H}_o^{-1}\rangle\,
%|\hat{\p}_{\vk}|^2 \psi(\T_{\vk},\phi)\nonumber\\
%&+& \f{k^2}{4\kappa}\, \langle \hat{H}_o^{-\f{1}{2}}\,
%\hat{a}^4(\phi)\, \hat{H}_o^{-\f{1}{2}}\rangle\, |\T_{\vk}|^2
%\psi(\T_{\vk},\phi)\Big]  \label{qevo2}\ea
%
where, by construction, the expectation values of the background
geometry operators are taken in the given state $\Psi_o$ of the
background quantum geometry.

Note that, at a fundamental level, $\psi$ now evolves in a
probability amplitude $\Psi_o$ of background geometries $g_{ab}$,
rather than a fixed $g_{ab}$. But (\ref{qevo2}) implies that its
evolution is not sensitive to \emph{all} the details of the
fluctuations of this quantum geometry; it is sensitive to only two
`moments' $\langle \hat{H}_o^{-1}\rangle$ and $\langle
\hat{H}_o^{-\f{1}{2}}\, \hat{a}^4(\phi)\,
\hat{H}_o^{-\f{1}{2}}\rangle$. This is so even though (\ref{qevo2})
is an \emph{exact} consequence of (\ref{qevo1}) with no further
approximations. Furthermore, since the back-reaction of the
perturbation on $\Psi_o$ is neglected within the test field
approximation inherent to our truncation scheme, nothing is lost by
projecting (\ref{qevo1}) along $\Psi_o$ in arriving at
(\ref{qevo2}). \emph{That is, within the test field approximation,
(\ref{qevo2}) captures the full information about the evolution of
$\psi$ that is contained in the original equation (\ref{qevo1}).}

Next, recall that in the standard cosmology literature one regards
the quantum state $\psi(\T_{\vk}, \phi)$ of perturbations as
propagating on a \emph{classical} FLRW metric, specified by the
scale factor $a_{\cl}(\phi)$. In the Schr\"odinger picture now under
consideration, the evolution is given by
\be i\hbar\partial_\phi \psi(\T_{\vk}, \phi) = \f{1}{2}\, \sum_{\vk}
4\kappa (\pphi^{-1})\, |\hat{\p}_{\vk}|^2
\psi(\T_{\vk},\phi)%\nonumber\\\,
+ \,\f{k^2}{4\kappa}\, [\pphi^{-1} a^4_{\cl}(\phi)]\,
|\hat{\T}_{\vk}|^2 \psi(\T_{\vk},\phi)  \label{qevo3}\ee
%
%\be \label{qevo3} i\hbar\partial_\phi \psi(\T_{\vk}, \phi) =
%\f{1}{2}\, \f{\ell^3}{(2\pi)^3}\, \int\! d^3k\,\,\Big[\, 4\kappa
%(\pphi^{-1})\, |\hat{\p}_{\vk}|^2 \psi(\T_{\vk},\phi)\, +\,
%\f{1}{4\kappa}\, [\pphi^{-1} a^4_{\cl}(\phi)]\, |\hat{\T}_{\vk}|^2
%\psi(\T_{\vk},\phi)\, \Big]\, , \ee
%
where $\pphi$ can also be expressed using geometric variables using
the constraint equation satisfied by the background. Comparing
(\ref{qevo3}) with (\ref{qevo2}), we find that the evolution of the
test perturbation $\hat{\T}_{\vk}$ on the \emph{quantum} background
geometry given by $\Psi_o(\nu,\phi)$ is indistinguishable from that of
a test perturbation propagating on a \emph{smooth} FLRW
background
\be \tilde{g}_{ab} dx^a dx^b \equiv d\tilde{s}^2 = -
(\tilde{p}_{(\phi)})^{-2}\, \ell^6\, \tilde{a}^6(\phi)\, d\phi^2 +
\tilde{a}(\phi)^2\, d{\x}^2 \ee
where
\be (\tilde{p}_{(\phi)})^{-1}  = \langle \hat{H}_o^{-1}\rangle
\quad\quad {\rm and} \quad\quad \tilde{a}^4 = \f{\langle
\hat{H}_o^{-\f{1}{2}}\, \hat{a}^4(\phi)\,
\hat{H}_o^{-\f{1}{2}}\rangle}{\langle \hat{H}_o^{-1}\rangle}\, .
\ee
Again, this is an \emph{exact} equivalence between our truncated LQG
and the theory of quantum fields on a smooth FLRW geometry
determined by $\tilde{a}$ and $\tilde{p}_{(\phi)}$.

At a technical level, the existence of such a simple relation is
quite surprising at first. But this result should \emph{not} be
interpreted to mean that the standard quantum theory of
perturbations on a classical FLRW solution to Einstein's equations
holds in the Planck regime. \emph{It does not!} For, the
$\tilde{a}(\phi)$ seen by $\hat{\T}_{\vk}$ is \emph{very} different
from the $a_{\cl}(\phi)$ of a classical solution; in particular,
$(\tilde{a}(\phi), \tilde{p}_{(\phi)})$ do \emph{not} satisfy
Einstein's equations. Indeed, their expressions involve $\hbar$;
although $\tilde{g}_{ab}$ is smooth, it incorporates quantum
corrections which are so large in the Planck regime that they tame
the big bang singularity. Furthermore, \emph{the pair does not even
satisfy the effective equations of LQC} which track the peak of the
wave function $\Psi_o(\nu,\phi)$ of the background geometry. Certain
aspects of quantum \emph{fluctuations} inherent in
$\Psi_o(\nu,\phi)$ are absorbed in these tilde fields. Thus,
$\tilde{g}_{ab}$ may be thought of as a \emph{dressed effective
geometry} that is relevant for propagation of linear perturbations
on the full quantum (background) geometry determined by $\Psi_o(\nu,
\phi)$. In retrospect, from what we know in other areas of physics,
such a result is not entirely unexpected. For example, light
propagating in a medium interacts with its atoms but the net effect
of these interactions can be encoded just in a few parameters such
as the refractive index of the medium. In our case, the `medium' is
the quantum geometry and the tilde variables $\t{a}, \t{p}_{(\phi)}$
encode the interaction between this `medium' and the perturbations.
Only two parameters suffice simply because the background quantum
geometry is homogeneous and isotropic. Already in the anisotropic
Bianchi models, (an extension of the discussion of \cite{dlt}
implies that) one would need additional parameters characterizing
the dressed, effective anisotropies. Finally, the encoding is rather
sophisticated: prior to the calculation, it would have been
impossible to guess the precise `moments' of the fluctuations of
geometry that are to
capture this interaction.\\

\emph{Remark:} Several equations in this sub-section closely
resemble those in \cite{akl}. However the conceptual under-pinning
is quite different. The discussion in \cite{akl} began by assuming
that quantum dynamics can be obtained by imposing the constraint
analogous to $[\h{\mathbb{S}}_o + \h{\mathbb{S}}^\prime_2]\,
\Psi_o\otimes\psi =0$. This `quantum constraint' was then expanded
and sub-leading terms were discarded to arrive at an equation
analogous to (\ref{qevo1}). Consequently it was not appreciated
that, within the truncation scheme, (\ref{qevo2}) carries the full
information about the quantum evolution of $\psi$. As discussed in
section \ref{s3.4} (see also Appendix A), a more careful examination
has revealed that the strategy of imposing a quantum constraint
cannot be justified. Therefore, we adopted a different route here.
We mimicked the strategy from the classical theory: Just as
$X^\alpha_{\rm Dyn}$ provides a lift of the dynamical trajectories
on $\ps_o$ to $\ps_{o}\times \ps_{1}$, we lifted the `quantum
dynamical trajectories' $\Psi_o$ on $\H^o_{\rm phy}$ to
$\Psi_o\otimes \psi$ on $\H^o_{\rm phy} \otimes \H^1$. This route is
also more direct in that we did not have to discard any terms to
arrive at (\ref{qevo1}).\\

Finally, let us translate this result using conformal time $\eta$
commonly used in the literature on cosmological perturbations. The
dressed, effective metric can be written as:
\be \tilde{g}_{ab} dx^a dx^b\, \equiv\, d\tilde{s}^2 \, = \,
\tilde{a}^2(\phi)\, (-d \tilde{\eta}^2 + d\x^2) \ee
where
\be d\tilde{\eta} = [\ell^3 \tilde{a}^2(\phi)]\,
\t{p}_{(\phi)}^{-1}\, d\phi \, .\ee
Therefore, in the truncated theory, the exact evolution equation for
the quantum perturbation $\hat{\T}_{\vk}$ on the background quantum
geometry is given by
\be \label{Teqn} \hat{\T}_{\vk}^{\prime\prime} + 2
\f{\tilde{a}^\prime}{\tilde{a}}\, \hat{\T}_{\vk}^\prime + k^2
\hat{\T}_{\vk} = 0\ee
where the prime now denotes a derivative with respect to
$\tilde\eta$. This mathematical equivalence simplifies both
conceptual and technical aspects of our analysis considerably
because the well-developed techniques from quantum field theory in
curved space-times can now be readily imported into the quantum
field theory of perturbations $\psi$ on quantum geometries $\Psi_o$.
In section \ref{s6}, we will use this strategy to define in detail
the quantum states for perturbations and composite operators that
are needed to complete the quantum theory.\\

\emph{Remarks:}\\
1. We can make a further simplification through a `mean field'
approximation in which the fluctuations are ignored. More precisely,
let us first recall \cite{aps3,acs,asrev} that even in the Planck
regime the state $\Psi_o(\nu,\phi)$ is sharply peaked on an
effective geometry
\be \bar{g}_{ab}dx^a dx^b\, \equiv\, d\bar{s}^2 \,=\, \bar{a}^6\,
\f{\ell^6}{\pphi^2}\, d\phi^2 \, +\, \bar{a}^2(\phi)\, d\x^2 \ee
which agrees with the general relativity solution (for the same
value of $\pphi$) for large $a(\phi)$ but has a in-built bounce at
$\bar{a}(\phi)^6 = \pphi^2/(2\ell^6\rho_{\rm max})$, with $\rho_{\rm
max} = {3}/(8 \pi G \gamma^2 \lambda^2) \approx 0.41 \rho_{\rm Pl}$.
In terms of our quantum geometry state $\Psi_o$, the scale factor is
given just by the expectation value $\bar{a}(\phi) = \langle
\hat{a}(\phi)\rangle$ in this state. Suppose we ignore quantum
fluctuations, i.e., \emph{use a mean field approximation} in which
the expectation values of powers of $\hat{a}$ and $\hat{H}_o$ are
replaced by the same powers of their expectation values. In this
approximation, the quantum perturbation $\hat{\T}_{\vk}$ would seem
to propagate on the effective geometry determined by the pair
$(\bar{a}(\phi), \pphi)$. Now, one knows that there exist background
quantum geometries $\Psi_o$ which are \emph{very} sharply peaked on
this effective geometry even in the Planck regime. If one uses such
a $\Psi_o$, the mean field approximation is excellent for studying
the propagation of perturbations on quantum geometries under
consideration.%
\footnote{In numerical simulations of the evolution of the quantum
state $\psi$ of perturbations, for example, if $\Psi_o$ is chosen
appropriately, the numerical errors would be much higher than those
introduced by the mean field approximation.}
The exact evolution of $\hat{\T}_{\vk}$, on the other hand, sees the
more sophisticated, `dressed' effective geometry determined by
$(\tilde{a}, \tilde{p}_{(\phi)})$.

2. If one were to use the mean field approximation, the quantum
perturbations would satisfy (\ref{Teqn}) with the tilde quantities
replaced by the barred quantities that refer to the effective LQC
solutions. At first sight, it may therefore seem one could have
arrived at these equations simply by perturbing the effective
equations of LQC. This could have been a viable interpretation had
there been a clear set of effective equations in full LQG to perturb
around backgrounds satisfying the LQC solutions. But as emphasized
in sections \ref{s1} and {\ref{s2}:\, i) we do not yet have
effective equations in full LQG, and, ii) if one were to adopt the
naive strategy of considering a set of linearized equations in
general relativity and simply replacing the background solution to
Einstein's equations by a solution to the effective equations, one
faces a very large ambiguity in the choice of equations with which
to begin and, furthermore, the set of final equations need not be
internally consistent when the background does not satisfy
Einstein's equations. Our procedure is free of these drawbacks
because we first constructed the quantum theory and arrived at Eq.
(\ref{Teqn}) by showing an exact equivalence of fields $\h{\Q},
\h{\T}$ propagating on the quantum geometry $\Psi_o$ and those
propagating on the geometry determined by $\t{g}_{ab}$. Nowhere did
this procedure use effective equations of LQC.

\section{Hilbert spaces of states of $\hat{\T}_{\vk}$}
\label{s5}

In this section we will construct the Hilbert space of quantum
states of gauge invariant perturbations following two different but
complementary avenues. The first is geared to mathematical
physicists and well adapted to the Hamiltonian framework used in our
classical considerations. The second follows the route that is more
often taken in analyzing cosmological perturbations. We show that
the two are equivalent. Therefore, to ensure conceptual continuity
and coherence, one can start with quantization given in section
\ref{s5.1} and then use the framework presented in \ref{s5.2} which
is better adapted to regularization, renormalization and numerical
simulations. We also provide the explicit expression of the 2-point
function which makes it manifest that the group of
space-translations continues to be a symmetry in the quantum theory.

\subsection{The Weyl algebra and its representations}
\label{s5.1}

The classical phase space $\Gamma_{\T}$ for the tensor modes is
spanned by canonically conjugate pairs $(\T_{\vk}, \p_{\vk})$. The
corresponding operators $\hat{\T}_{\vk}, \hat{\p}_{\vk}$ satisfy the
canonical commutation relations and generate the Heisenberg algebra
in the quantum theory. For technical simplicity, it is convenient to
exponentiate them to obtain the Weyl algebra $\W$ whose
representations provide the required Hilbert spaces of quantum
states.

One begins with the observation that, with each vector $(\lambda,\mu)
\in \Gamma_{\T}$, one can associate a natural linear combination of
smeared configuration and momentum operators:
\ba \hat{F}(\lambda,\mu) := \Omega((\hat{\T},
\hat{\p}),\,\,(\lambda,\mu)) &=& \f{1}{\Vzero}\, \sum_{\vk}
\mu^\star_{\vk} \hat{\T}_{\vk} \, -\, \lambda^\star_{\vk}
\hat{\p}_{\vk}\nonumber\\
&=& \int_{\man} d^3\x\,\, (\mu(\x) \hat{\T}(\x) - \lambda(\x)
\hat{\p}(\x))\, . \ea
%\ba \hat{F}(\lambda,\mu) := \Omega((\hat{\T},
%\hat{\p}),\,\,(\lambda,\mu)) &=& \int\! d^3k\,\, (\mu^\star_{\vk}
%\hat{\T}_{\vk} \, -\, \lambda^\star_{\vk}
%\hat{\p}_{\vk})\nonumber\\
%&=& \int_{\man} d^3\x\,\, (\mu(\x) \hat{\T}(\x) - \lambda(\x)
%\hat{\p}(\x))\, . \ea
%
The Weyl operators $\hat{W}(\lambda, \mu)$ are their exponentials:
\be \hat{W}(\lambda,\mu) := e^{\f{i}{\hbar}\,
\hat{F}(\lambda,\mu)}\, . \ee
It is more convenient to work with these exponentials for two
reasons. First, while the field operators $\hat{F}$ are unbounded,
the $\hat{W}$ are unitary and hence bounded operators in any
representation. Therefore one avoids the awkward issues of
specifying operator domains. Second, the vector space generated by
finite linear combinations of the Weyl operators is automatically
closed under the Hermitian-conjugation operation
$\hat{W}^\dag(\lambda,\mu) = \hat{W}(-\lambda,\, -\mu)$, and, more
importantly, under the product:
\be \hat{W}(\lambda_1,\mu_1)\, \hat{W}(\lambda_2,\mu_2) =
e^{\f{1}{i\hbar}\, \int d^3\x\, (\lambda_1\mu_2 -
\mu_1\lambda_2)}\,\, \hat{W}(\lambda_1+\lambda_2,\mu_1+\mu_2)\, .
\ee
Thus, the vector space has the structure of a $\star$-algebra. This
is the Weyl algebra $\W$. (It can be easily endowed the structure of
a $C^\star$ algebra but this will not be necessary for our
purposes.)

To find the representations of $\W$, it is simplest to use the
standard Gel'fand, Naimark, Segal (GNS) construction \cite{gns}:
Given a  positive linear function (PLF) on $\W$, the construction
provides an explicit Hilbert space $\H_1$ and a representation of
elements of $\W$ by concrete operators on that $\H_1$. This
representation is cyclic: every state in $\H_1$ arises from the
action of operators representing elements of $\W$ on a `vacuum'. In
this representation, the PLF turns out to be just the vacuum
expectation value functional. A natural strategy for linear fields
is to first find a complex structure on the phase space that is
compatible with the symplectic structure thereon and then use the
Hermitian inner product provided by the resulting K\"ahler structure
to define the required PLF on $\W$ \cite{am,waldbook}. We will follow
this conceptual strategy but in a manner that retains close contact
with the cosmology literature. Therefore, prior knowledge of the
complex and K\"ahler structures will not be necessary to follow the
construction.

In linear field theories in Minkowski space, one narrows the
selection of the positive linear functional by requiring that it
(and hence the vacuum state) be Poincar\'e invariant. In the present
case, it is natural to require that the PLF be invariant under the
3-dimensional translational symmetry of the background geometry.
Such PLFs can be constructed as follows. Choose a set of complex
coefficients $(e_k, f_k)$ (with $k\ge 0$) such that
\be \label{data} e_k f^\star_{k} - e^\star_k f_k  \equiv 2i \, {\rm
Im}\, (e_k f^\star_k) = i \quad {\hbox{\rm for all}}\,  k\ee
(For a massless scalar field in Minkowski space, $e_k = e^{-i\omega
t_o}/\sqrt{2\omega}$ and $f_k = (-i\omega\, e^{-i\omega
t_o})/\sqrt{2\omega}$ with $\omega = |\vk|$.) Then, one can extract
the `positive frequency' part $a_{\vk}$ of any vector $(\lambda,\mu
) \in \Gamma_{\T}$ as follows:
\be a_{\vk}\, :=\, -i\, (f^{\star}_{k} \lambda_{\vk} - e^\star_{k}
\mu_{\vk}) \ee
so that
\be \lambda_{\vk} = e_k a_{\vk} + e_k^{\star} a^{\star}_{\vk}, \quad
{\rm and} \quad \mu_{\vk} = f_k a_{\vk} + f^{\star}_{k}
a^{\star}_{\vk}\, . \ee
(Note that while $\lambda,\, \mu$ satisfy the `reality condition'
$\lambda_{\vk}^\star = \lambda_{-\vk}, \,\, \mu_{\vk}^\star =
\mu_{-\vk}$, the `positive frequency parts' don't: $a_{\vk}^\star
\not= a_{-\vk}$.) The required PLF is then simply
\be \label{plf} \langle \hat{W}(\lambda, \mu) \rangle =
e^{-\f{1}{2\hbar}\, \f{1}{\Vzero}\, \sum |a_{\vk}|^2} \ee
%
%\be \label{plf} \langle \hat{W}(\lambda, \mu) \rangle =
%e^{-\f{1}{2\hbar}\, \f{1}{\ell^3}\, \sint\!d^3k\,\, |a_{\vk}|^2} \ee
%
Under the action of a translation $\x \to \x+\vec{v}$ on $\man$, we
have: $\hat{\T}(\x) \to \hat{\T}(\x+ \vec{v})$,\, and\, $\hat\p(\x)
\to \hat\p(\x+ \vec{v})$,\,\, whence $\hat{W}(\lambda(\x), \mu(\x))
\to \hat{W}(\lambda(\x -\vec{v}), \mu(\x-\vec{v}))$, or, in the
momentum space, $\lambda_{\vk} \to e^{-i{\vk}\cdot
\vec{v}}\lambda_{\vk},\,\, \mu_{\vk} \to e^{-i{\vk}\cdot
\vec{v}}\mu_{\vk}$. This trivially implies $a_{\vk} \to
e^{-i\vk\cdot \vec{v}} a_{\vk}$. Hence the PLF (\ref{plf}) is left
invariant and the translation is represented by a unitary
transformation on the GNS Hilbert space $\H_1$ under which the GNS
vacuum is invariant. Thus, each choice of coefficients $(e_{k},\,
f_{k})$ satisfying (\ref{data}) leads to a representation of the
Weyl algebra in which the (GNS) vacuum is invariant under
translations.

\subsection{Space-time description} \label{s5.2}

At this stage it is convenient to make the construction more
explicit using the familiar expansions in terms of the creation and
annihilation operators. This entails going to the Heisenberg
picture. In terms of the conformal time $\t\eta$ of the dressed
effective metric of section \ref{s4.1}, the space-time operator is
represented as $\h{\T}(\x,\t\eta)$. It satisfies the field equation
and the canonical commutation relations, and is self-adjoint. These
properties can be neatly captured by an expansion of the form
\be \label{expansionT}\h{\T}(\x, \t\eta) = \f{1}{\Vzero}\,
\sum_{\vk} \big(e_k (\t\eta) \h{A}_{\vk} + e^\star_k (\t\eta)
{\h{A}_{-\vk}}^{\dag} \big)\, e^{i\vk\cdot\x}\, .\ee
%
%\be \label{expansionT}\h{\T}(\x, \t\eta) = \f{1}{(2\pi)^3}\,
%\int\!d^3k\,\, \big(e_k (\t\eta) \h{A}_{\vk} + e^\star_k (\t\eta)
%{\h{A}_{-\vk}}^{\dag} \big)\, e^{i\vk\cdot\x}\ee
%
We require that $e_k(\t\eta)$ should satisfy (\ref{Teqn}):
\be \label{eome} e_k^{\prime\prime}(\t\eta) +
2\f{{\t{a}}^\prime}{\t{a}}\, e_{k}^\prime(\t\eta) + k^2 e_k(\t\eta)
= 0\, , \ee
so that $\h{\T}(\x,\t\eta)$ satisfies the desired equation of
motion. (In relation to the more familiar expansion in Minkowski
space-time, $e_{k}(\t\eta)$ now plays the role of the \emph{positive
frequency} basis functions $e^{-i\omega t}/\sqrt{2\omega}$.) Next,
for each $\vk$, the space of solutions to (\ref{eome}) is two
dimensional and one chooses a complex solution $e_k(\t\eta)$
satisfying the normalization condition:
\be \label{normalized}\f{\t{a}^2}{4\kappa}\, \big(e_k(\t\eta)\,
e^{\prime\star}_k(\t\eta)\, -\, e^\star_k(\eta)\, e^\prime_k(\eta)
\big) = i \, .\ee
This condition needs to be imposed only at some initial instant of
time $\t\eta_o$; Eq (\ref{eome}) then guarantees that it is then
automatically satisfied for all $\t\eta$. With this normalization,
if the time independent operators $\h{A}_{\vk}$ and
$\h{A}^\dag_{\vk}$ satisfy the commutation relations
\be \label{comrel} [\h{A}_{\vk},\,\, {\h{A}_{\vk^\prime}}^\dag ] =
\hbar\, \ell^3\, \delta_{\vk, \vk^\prime}\, , \ee
then
\be \h\T (\x, \t\eta)\quad {\rm and} \quad  \h\p(\x, \t\eta) =
\f{\t{a}^2}{4\kappa}\,\, \f{1}{\Vzero}\, \sum_{\vk} \big(e^\prime_k
(\t\eta) \h{A}_{\vk} + e^{\star\prime}_k (\t\eta)
{\h{A}_{-\vk}}^{\dag} \big)\, e^{i\vk\cdot\x}\, .\ee
satisfy the required canonical commutation relations at any fixed
value of $\t\eta$. In view of their properties, one can interpret
$\h{A}_{\vk}$ as annihilation operators, define the vacuum $|
0\rangle$ as the state annihilated by all $\h{A}_{\vk}$, and
generate the Fock space $\H_1$ by repeatedly acting on the vacuum by
creation operators $\h{A}^\dag_{\vk}$. It is straightforward to
calculate the vacuum expectation value of the Weyl operator
\be \h{W}(\lambda, \mu)|_{\t\eta_o} = e^{\f{i}{\hbar\Vzero}\,
\sum_{\vk} \mu^\star_{\vk} \hat{\T}_{\vk} \, -\, \lambda^\star_{\vk}
\hat{\p}_{\vk}}\ee
at any conformal time $\t\eta_o$. It is given by (\ref{plf}) where
$e_k
= e_k(\t\eta_o)$ and $f_k = (a^2/4\kappa)\, e^\prime_k(\t\eta_o)$.\\
Thus, our second description of the Hilbert space $\H_1$ adapted to
the covariant space-time picture, is completely equivalent to the
first description, adapted to the Weyl algebra that was constructed
starting from the phase space. The first description serves to bring
out the conceptual structure of the quantum theory that emerges from
the phase space description of section \ref{s3} while the second is
better adapted to calculations, e.g. regularization of the
stress-energy tensor discussed in section \ref{s6}.

To summarize, for each choice of solutions $e_k(\t\eta)$ satisfying
the normalization condition (\ref{normalized}) we obtain a vacuum
state and hence a Fock representation of the canonical commutation
relations (or, of the Weyl algebra). Since these representations are
completely characterized by their 2-point functions, it is
instructive to write them out explicitly. Using the expansion
(\ref{expansionT}) of $\h\T(\x,\t\eta)$, we obtain:
\be \langle 0\mid \h\T(\x_1, \t\eta_1)\, \h\T(\x_2, \t\eta_2) \mid
0\rangle = \f{\hbar}{\Vzero}\, \sum_{\vk} e^{i\vk\cdot(\x_1-\x_2)}\,
e_k (\t\eta_1) e^\star_k(\t\eta_2)\, . \ee
By inspection the 2-point function is invariant under the action of
space-translations. This is an independent proof of the
translational invariance of the vacuum state, now geared to the
cosmology literature.\\

\emph{Remark}: While each choice of the family of solutions
$e_{k}(\t\eta)$ to (\ref{eome}) satisfying the normalization
condition (\ref{normalized}) determines a vacuum state (and the
associated Hilbert space $\H_1$ of states), this is a many to one
map: The families $e_{k}(\t{\eta})$ and $e^{i\theta_k}\,
e_k(\t{\eta})$ that differ just by a $k$-dependent phase factor
determine the same vacuum. There is a 1-1 correspondence between the
equivalence classes $\{e_{k}(\t\eta)\}$ of families that differ by
such phase factors and complex structures $J$ on the phase space
$\Gamma_{\T}$ which are compatible with the symplectic structure and
are invariant under the action of space-translations. Thus there is
a 1-1 correspondence between the vacua $|0\rangle$ we have
constructed and complex structures on
$\Gamma_{\T}$ satisfying the two properties listed above.\\

A natural question about these representations of the Weyl or the
Heisenberg algebra is now the following: Does a change of the
complex structure $J$ ---i.e., the choice of `generalized positive
frequency solutions' $\{e_k(\t\eta)\}$--- always result in unitarily
equivalent representations? As is well-known, in general the answer
is in the negative. In the terminology used in the cosmology
literature, in general the vacuum state selected by any one complex
structure may contain an infinite number of particles corresponding
to another complex structure. A priori this would be a key obstacle
to extracting physics because different choices would in general
lead to very different predictions. Furthermore, in a general
representation so constructed, there is no natural prescription to
regulate products of operator-valued distributions, e.g.
$\h{\T}^{2}(x,\t{\eta})$, and hence to define basic physical
operators such as the Hamiltonian of (\ref{qevo2}).

Fortunately, as we will see in section \ref{s6}, both these problems
can be resolved in one stroke by imposing certain regularity
requirements on the basis functions $\{e_k(\t\eta)\}$. Then, the
representations of the Weyl algebra that result from equivalence
classes $\{e_k(\t\eta)\}$ of any of the regular basis functions
---or `regular' complex structures $J$--- will turn out to be
unitarily equivalent. In this sense there is a unique class of
unitarily equivalent representations and one can work with a unique
Hilbert space $\H_1$. Therefore the more general framework of
algebraic quantum field theory is not essential in the cosmological
context under consideration.%
\footnote{See also \cite{iberian,madrid2,madrid3} where one arrives
at a rigorous uniqueness result using the form (\ref{eom}) of
equations of motion. However, the `vacua' they are led to consider
are not necessarily of 4th adiabatic order whence it would be
difficult to regularize and renormalize physically interesting
composite operators, such as the energy density, in that framework.
It would be interesting to see in detail the precise relation
between that approach and the adiabatic treatment pursued here and
in much of the cosmological literature.}

All the regular translationally invariant vacua and states
containing a finite number of excitations over any of them belong to
the Hilbert space $\H_1$. However, the translationally invariant
vacua span an \emph{infinite dimensional} subspace of $\H_1$ and
none of them is preferred from a full space-time perspective. Thus,
$\H_1$ will not admit a `canonical' vacuum we are used to in
Minkowski or (strictly) stationary space-times.\\

\section{Regularity conditions on states and operators}
\label{s6}

In this section we will first summarize the notion of regularity
conditions on states $\psi$ of quantum perturbations and then use it
to regulate products of operator-valued distributions, such as the
ones that appear in the quantum stress-energy tensor or Hamiltonian.
There exist several methods of regularization. We will work with the
adiabatic scheme because it is particularly suitable to perform
explicit computations, including numerical implementations, and can
be directly extended to our quantum field theory on \emph{quantum}
FLRW geometries. In this paper of course we will focus on the
$k\!=\!0,\, \Lambda\!=\!0$ case but our considerations will extend
to other contexts such as the $k\!=\! 1$ FLRW case where the
underlying isometries make a mode decomposition naturally available.

Much of the discussion in the first two sub-sections is taken from
the rich literature on adiabatic regularization in cosmology (see,
e.g., \cite{anderson-parker,parker-fulling74,sf-book,parker-book}).
But there are two new elements as well: i) the specific formulation
of the adiabatic condition (which is succinct and yet clarifies some
subtleties); and ii) the discussion of the regularized Hamiltonian
operator $\h{H}_1$.

\subsection{The adiabatic condition}
\label{s6.1}

As explained in Section \ref{s5.2}, a choice of a basis of
`generalized positive frequency' solutions $e_{k}(\t\eta)$
satisfying the normalization condition (\ref{normalized}) determines
a vacuum state, $|0\rangle$, from which a Fock space $\H_1$ can be
constructed. Since each of these (complex) basis vectors
$e_{k}(\t\eta)$ satisfies the second order, linear, ordinary
differential equation (\ref{eome}), any two  bases $e_{k}(\t\eta)$
and ${\ub{e}}_{k} (\t\eta)$ are related simply by
\cite{parker66,parker69}
\be \label{bogrel}  e_{k}(\t\eta)= \alpha_k \, {\ub{e}}_{k}(\t\eta)+
\beta_k \, {\ub{e}}_{k}^{\star}(\t\eta) \, , \ee
where the \emph{time-independent} Bogoluibov coefficients $\alpha_k$
and $\beta_k$ satisfy the relation $|\alpha_k|^2-|\beta_k|^2=1$.
Substituting this equation in the expression (\ref{expansionT}) of
the field operator $\h{\T}(\x, \t\eta)$, we find a linear relation
between the creation and annihilation operator associated with the
two families

\be \label{bogrelop} \hat{{\ub{A}}}_{\vec{k}}= \alpha_k \, \hat
A_{\vec{k}}+ \beta^{\star}_k \, \hat A^{\dagger}_{-\vec{k}} \, . \ee
This relation shows that, as long as the $\alpha_k$ coefficient are
not trivial, i.e. $\alpha_k \not= e^{i\theta_k}$ for all $k$, the
associated vacua $|0\rangle$ and $|{\ub{0}}\rangle$ are distinct.
The number of `under-barred' quanta with momentum $\vk$ that the
state $|0\rangle$ contains is given by
\be \label{number} \langle 0 | \hat{{\ub{N}}}_{\vec{k}}| 0 \rangle
:= \langle 0 | (\hbar\ell^3)^{-1}\hat{{\ub{A}}}_{\vec{k}}
\hat{{\ub{A}}}^{\dagger}_{\vec{k}} | 0 \rangle=  \, |\beta_k|^2
%(2\pi)^3 \delta^3(0)
\, ,\ee
where we have used (\ref{bogrelop}) and the commutation relation
(\ref{comrel}) in the last step. The right hand side provides the
expected number of `under-barred exitations/particles' with momentum
$\vk$ in the vacuum $|0\rangle$. Therefore, at the power counting
level, it follows that if $|\beta_k|^2$ does not fall-off faster
than $k^{-3}$ when $k\to\infty$, the total number of `under-barred
quanta' in the vacuum $|0\rangle$ would diverge. In this case,
$|0\rangle$ would not belong to the under-barred Fock space; the two
representations of the Heisenberg or Weyl algebra would be unitarily
inequivalent.

This inequivalence is related to the large $k$ limit and, as
indicated in section \ref{s5.2}, can arise because so far there is
no restriction on the ultraviolet behavior of the basis states
$e_{k}(\t\eta)$. The physical idea behind the appropriate
restriction becomes clearer if one works with the variable
$\chi_k(\t\eta):= (\t{a}(\t\eta)/2\sqrt{\kappa})\, e_{k}(\t\eta)$,
for which (\ref{eome}) becomes
\be \label{leq} \chi_k''(\t\eta)+\left(k^2-
\frac{{\t{a}}''(\t\eta)}{\t{a}(\t\eta)}\right) \, \chi_k(\t\eta)=0
\, . \ee
Note that (\ref{leq}) reduces to the equation satisfied by the
standard basis functions in Minkowski space if ${\t{a}}''/\t{a} =0$.
Now, $6{\t{a}}''/{\t{a}}^3$ is just the scalar curvature of
$\t{g}_{ab}$ and introduces a physical length scale $L(\t{\eta})$
into the problem. The form of (\ref{leq}) suggests that for modes
with large momentum, i.e., with $k/\t{a} \gg 1/L$, curvature would
have negligible effect and they would evolve almost as if they were
in Minkowski space. Therefore, it is natural to impose the following
regularity condition on the choice of basis functions in the
ultraviolet limit: for $k/\t{a} \gg 1/L$,\, $\chi_{k}(\t\eta)$
should approach the canonical Minkowski space \emph{positive
frequency} solutions $e^{i k t}/\sqrt{2 k}$ at an appropriate rate.
(In the terminology of section \ref{s5.2}, we would then be
restricted to a preferred family of complex structures all of which
have the same ultraviolet behavior as the canonical complex
structure in Minkowski space-time.) This is the crux of the idea
behind adiabatic condition.

To make this idea precise, we have to sharpen the required rate of
approach. For this, one first introduces a set of specific
`generalized WKB' solutions to (\ref{leq}) that approach the
Minkowski space positive frequency modes as $k \to \infty$ in a
controlled fashion. In a second step, one requires that the
permissible basis functions $\chi_k(\t\eta)$ ---which are exact
solutions to (\ref{leq})--- should approach the WKB solutions to the
desired order.

The generalized WKB solutions $\chi_{k}^{(N)}(\t\eta)$ of order $N$
are given by \cite{parker-fulling74}:
\footnote{For this prescription to be well-defined,
$W_k^{(N)}(\t{\eta})$ must be non-negative. For any given smooth
$\t{a}(\t\eta)$, this can be ensured by going to high enough $k$.
Since it is only the behavior of $W_k^{(N)}(\t{\eta})$ in the $k\to
\infty$ that matters for the adiabiticity considerations, if
$W_k^{(N)}$ were to become negative in some $k$-range, one can just
suitably modify its form for low $k$.}
\be \label{approxsol} \chi_{k}^{(N)}(\t\eta)=\frac{1}{\sqrt{2 \,
W^{(N)}_k(\t\eta)}} \, e^{-i \int^{\t\eta} W^{(N)}_k(\tau) \,
d{\tau} } \, \ee
where $W^{(N)}_k(\t\eta)=W_0+ \, W_1+...+ \, W_N$, with
\ba W_0&=& k\, ; \quad W_2= -\frac{1}{2\, k} \frac{{\t{a}}''}{\t{a}}
\, ; \quad  W_4=\frac{2 {\t{a}}'' {\t{a}}'^2 - 2 {\t{a}}''^2 \t{a} -
2 \t{a} {\t{a}}' \t{a}^{'''} + \t{a}^2 \t{a}^{''''}}
{8 k^3 \t{a}^3} \, ; \quad ... \nonumber \\
{\rm and}  & & W_i=0 \quad {\hbox{\rm if $i$ is odd}} \, . \ea
Note that, because the lower bound on the integral in the phase
factor is not fixed, $\chi_{k}^{(N)}$ is well-defined only up to an
overall phase (which is time independent but can depend on $k$).
Each $\chi_{k}^{(N)}(\t\eta)$ is an approximate solution to
(\ref{leq}) in the sense that, when we operate on it by the operator
on the left side of (\ref{leq}), the result does not vanish but is
given by terms of the order $\mathcal{O}(\t{a}/kL_{N+2})^{N+1/2}$
where the length scale $L_N$ is dictated by $\t{a}$ and its
$\t\eta$-derivatives to order $N$. The leading order term of
(\ref{approxsol}) corresponds to the positive frequency solution in
Minkowski space and the rest of the terms are higher order
contributions that vanish at different rates when $(\t{a}/kL_N) \to
0$. Finally, this approximate solution can also be regarded as an
expansion in the number of derivatives of the scale factor. This is
because, since the limit refers to $(\t{a}/kL_N)$, one can either
keep $L_N$ fixed and let $\t{a}/k$ go to zero as we have done so
far, or keep $\t{a}/k$ fixed and let $1/L_N$ go to zero, which
corresponds to letting the expansion rate go to zero, i.e,
considering adiabatic expansion. This why, although this method is
primarily concerned with ultraviolet issues, it is
referred to as  `adiabatic regularization'.%
\footnote{Chronologically, the term {\em adiabatic} originated from
the fact that the first application of this method ---introduced by
Parker in \cite{parker66}--- was to the problem of regularizing the
number operator in an expanding universe, and the condition that the
particle number be an adiabatic invariant was used as a fundamental
requirement in the construction.}

We are now ready to state the `adiabatic condition' that imposes the
desired ultraviolet regularity on the basis functions: \emph{In the
mode expansion (\ref{expansionT}), choose only those solutions
$e_{k}(\t\eta)= 2\sqrt{\kappa}\chi_{k}(\t\eta)/{\t{a}}(\t\eta)$\,\,
to (\ref{eome}) which agree with  $\chi^{(N)}_{k}(\t\eta)$ up to
terms of order $(\t{a}/kL_{N})^{N+1/2}$.} More precisely, we
require: $|\chi_{k}| = |\chi^{(N)}_{k}|\,\,\,
(1\,+\,\mathcal{O}((\t{a}/kL_{N+2}))^{N+2})$}, and the same relation
should hold if $\chi_k$ and $\chi^{(N)}_{k}$ are replaced by their
1st, 2nd, ... (N-1)th time derivatives. (Note that the absolute
value signs make these conditions well-defined in spite of the phase
ambiguity in $\chi^{(N)}_{k}$.) These bases $e_{k}(\t\eta)$ will be
referred to as $Nth$-order adiabatic solutions, the associated
vacuum states will be referred to as the \emph{$Nth$ order vacua},
and states obtained by operating on these vacua by (arbitrarily
large but finite) sums of products of creation operators as
$Nth$-order adiabatic states. From now on, we will restrict
ourselves to $Nth$-order adiabatic states and, for reasons explained
in section \ref{s6.2}, for most part we will set $N=4$. There is a
large body of literature on the notion of adiabatic states and their
properties. For further
details, see in particular \cite{parker-fulling74,sf-book,parker-book}.\\

This framework has two important features.
\begin{itemize}
\item The adiabatic condition is only an {\em asymptotic}
    restriction for large $kL/\t{a}$. Therefore, for any given
    $N$, there are infinitely many families of solutions
    $e_{k}(\t\eta)$ which satisfy it. Each of these bases
    defines an adiabatic vacuum and, if two bases are
    non-trivially related (i.e., if the Bogoluibov coefficients
    $\alpha_k$ are not pure phases for all $k$), the
    corresponding vacua are distinct. Thus, in striking contrast
    to the free field theory in Minkowski space, there is no
    preferred vacuum state, nor a canonical notion of
    `particles'. Any one vacuum appears as an `excited state
    with many particles' with respect to another vacuum.

\item However, if $N\ge 2$, all adiabatic vacua ---and hence all
    adiabatic states--- lie in the same Hilbert space $\H_1$.
    This is because if $N\ge 2$, then $\sum |\beta_{k}|^2
    <\infty$, whence any one adiabatic vacuum has only a finite
    number of particles relative to any other.

\end{itemize}

This completes the specification of the Hilbert $\H_1$ of
perturbations we began in section \ref{s5.2}. While we used the
framework originally developed for quantum field theory in classical
space-times, because we used the dressed effective metric
$\t{g}_{ab}$ for our background, it follows from section \ref{s4}
that $\H_1$ is the Hilbert space of quantum perturbations $\psi$
propagating on the quantum geometry $\Psi_o$.\\

\emph{Remark:} Had we chosen to work with spatial manifolds $\man$
with $\mathbb{R}^3$ topology rather than $\mathbb{T}^3$, Eq
(\ref{number}) would be replaced by
\be \label{number2} \langle 0 | \hat{\ub{N}}_{\vec{k}}| 0 \rangle :=
\langle 0 | \hbar^{-1}\hat{\ub{A}}_{\vec{k}}
\hat{{\ub{A}}}^{\dagger}_{\vec{k}} | 0 \rangle = \, |\beta_k|^2
(2\pi)^3 \delta^3(0) \, .\ee
This implies that $|\beta_k|^2$ is now the number \emph{density} of
the `under-barred exitations/particles' ---with momentum $\vk$--- in
the vacuum $|0\rangle$ \emph{per unit volume of $\man$ and per unit
co-moving volume in the momentum space}. The $\delta^3(0)$ (in the
momentum space) in (\ref{number2}) arises because of the infinite
spatial volume of $\man$. Thus, in addition to the potential
ultraviolet divergence that would occur if $|\beta_k|$ does not
fall-off appropriately for large $k$, we now also have an infrared
divergence. Note that this cannot be cured simply by putting an
infrared cut-off in the $\vk$ space: the \emph{total} number of
particles created with momenta $\vk$ within \emph{any finite range}
$\Delta_k$ also diverges because of the infinite spatial volume of
$\man$. In particular, this divergence persists for massive fields
as well; it arises  because we now have an infinite volume
\emph{and} a homogeneous background. This then implies that the Fock
representations of the Heisenberg or Weyl algebra associated with
any two bases $e_k(\t\eta)$ and ${\ub{e}}_k(\t\eta)$ are unitarily
inequivalent unless $\alpha_k$ is a pure phase for all $k$. But
physically this infinity is spurious. Therefore, for $\mathbb{R}^3$
topology, a notion of `physical equivalence' is more appropriate
than that of `unitary equivalence'. To introduce it, recall first
that if the topology is $\mathbb{R}^3$ one has to introduce a
fiducial cell $\mathcal{C}$ and restrict all integrations to it
already at the classical level. (From a physical perspective one can
choose the cell so that its physical volume is larger than the
volume of the observed universe.) Two Fock representations of
perturbations would be regarded as \emph{physically equivalent} if
the vacuum state associated with one contains a finite number of
`exitations/particles' with respect to the other within any region
contained in $\mathcal{C}$. Then, if we restrict ourselves to
adiabatic vacua of order $N\ge 2$, we are ensured of physical
equivalence. If we require $N\ge 4$, the expectation values of the
regularized stress-energy tensor would be well-defined distributions
and we can restrict the support of test functions in $\mathcal{C}$.

\subsection{Regularization of composite operators}
\label{s6.2}

In this subsection we summarize the necessary regularization
procedure to obtain physical information from the formal expressions
of composite operators on the Hilbert space $\H_1$. Adiabatic
regularity of the basis modes $e_k(\t\eta)$ will provide the
necessary control on the ultraviolet divergences in the expectation
values of composite operators, leading to state \emph{independent}
criteria to extract the physical, finite results while respecting
the underlying covariance of the theory.

Consider a formal operator $\h{O}(\x, \t{\eta})$ which is at least
quadratic in the field operator and its conjugate momentum, and is
factor ordered to be self-adjoint. Examples of direct interest are
the stress-energy tensor and the Hamiltonian. Consider the
expectation value of this operator with respect to an adiabatic
vacuum selected by a basis $\chi_k(\t\eta)$. Using the mode
decomposition (\ref{expansionT}) and the commutation relation
(\ref{comrel}), the vacuum  expectation value of
$\h{O}(\x,\t{\eta})$ can be expressed as a formal sum in the
momentum space, of the type

\be \langle 0| \h{O}(\vec{x},\t\eta)|0\rangle_{\rm formal} =
\f{1}{\ell^3}\, \sum_{\vk}\, O_k[\chi_k(\t\eta)] 	\,  \ee
where there is no $x$-dependence on the right side because
$|0\rangle$ is translationally invariant. Generically, the sum would
be ultraviolet divergent. In adiabatic regularization the physically
relevant, finite expression is obtained by subtracting, \emph{mode
by mode}, each term in the adiabatic expansion of
$O_k[\chi_k(\t\eta)] $ that contains at least one ultraviolet
divergent piece \cite{parker66,parker-fulling74} (see also
\cite{sf-book,parker-book}). Thus, if $O^{(N)}$ is the $Nth$
adiabatic order term in the expansion of the summand, if any one
part of $O^{(N)}$ is divergent, the entire term $O^{(N)}$ is
subtracted (including parts that have no ultraviolet divergences).
On the other hand, following the \emph{criterion of minimal
subtraction}, this procedure is applied only up to that order in the
adiabatic expansion at which the formal expression has divergent
pieces. Thus, if the $Mth$ adiabatic order term has no divergent
part, then nothing is subtracted at order $M$.

The most interesting example for us is the stress-energy tensor of
gauge invariant tensor perturbations since it plays the key role in
checking if the truncation scheme is self-consistent, i.e., whether
or not the back-reaction can be neglected. For classical fields
$\T(\x)$, the expression is given by
\be T_{ab}= \f{1}{4\kappa}\,\, \big[\t\nabla_a\T \, \t\nabla_b\T
-\frac{1}{2} \, \t{g}_{ab} \, \t{g}^{cd} \, \t\nabla_c\T \,
\t{\nabla}_d\T \big]\, .\ee
At the quantum level the stress-energy tensor is a composite
operator of dimension four, and in four space-time dimensions we
expect ultraviolet divergences up to fourth order in the co-moving
momentum $k$. Let us first consider energy density operator
$\hat{\rho} (\x, \t{\eta}) $. Given a basis $e_{k}(\t\eta) =
2\sqrt{k}\,\chi_{k}(\t\eta)/\t{a}(\t\eta)$ of fourth or higher
adiabatic order, the formal expression for the expectation value of
$\h{\rho}(\x,\t{\eta})$ in the associated adiabatic vacuum is
\ba \langle 0| \h\rho |0\rangle_{\rm formal}&:=&-\langle 0| \h T^0_0
|0\rangle_{\rm formal} =\frac{\hbar}{\ell^3 {\t{a}}^{4}(\t\eta)}
\sum_{\vk} \, \rho_k[\chi_k(\t\eta)] \nonumber \\ &=&
\frac{\hbar}{2\ell^3 {\t{a}}^4(\t\eta)} \sum_{\vk} \, |\chi'_k|^2+
\left(k^2+\frac{{\t{a}}'^2}{\t{a}^2} \right) |\chi_k|^2-2
\frac{\t{a}'}{\t{a}} {\rm Re}( \chi_k \chi'^{\star}_k ) \, . \ea
By using the adiabatic expansion of each $\chi_k$ in the above
summand, it is easy to see that all the ultraviolet divergences are
contained in terms of adiabatic order equal to and smaller than
four. The zeroth adiabatic order term produces a $\sum (1/k^3)$
divergence in co-moving momentum $k$; the second order term a $\sum
(1/k^2)$ one; and the fourth order term a $\sum (1/k)$ one.
Therefore, the subtraction terms $C^{\rho}(k,\t\eta)$ needed to
regularize the energy density are obtained from the terms of zeroth,
second and fourth adiabatic order in the expansion of the summand:
\be C^{\rho}(k,\t\eta)=
\rho_k^{(0)}+\rho_k^{(2)}+\rho_k^{(4)}=\frac{k}{2
}+\frac{{\t{a}}'^2}{4 {\t{a}}^2 k}+\frac{4 {\t{a}}'^2
{\t{a}}''+\t{a} {\t{a}}''^2-2 \t{a} {\t{a}}'\, {\t{a}}^{'''}}{16
{\t{a}}^3 k^3} \, . \ee
Thus, the vacuum expectation value of the renormalized energy
density is
\be \label{renden} \langle 0| \h\rho |0\rangle_{\rm
ren}=\frac{\hbar}{\ell^3 \t{a}(\t\eta)^4} \sum_{\vk}  \,
(\rho_k[\chi_k(\t\eta)] -C^{\rho}(k,\t\eta)) \, .\ee
Note that the subtraction terms $C^{\rho}$ are local in the
background geometry and, even more importantly, are {\em state
independent}. The expectation value in any fourth order state in
$\H_1$ is computed by the same procedure, using the \emph{same
subtraction terms}.

The only other independent degree of freedom in the stress-energy
tensor in a homogenous and isotropic background is the trace $T$ (or
the pressure $p=(\rho+T)/3$). The corresponding vacuum expectation
value can be calculated using results given in
\cite{anderson-parker}:
\be \label{renpre} \langle 0| \h T |0\rangle_{\rm
ren}=\frac{-\hbar}{\ell^3 {\t{a}}^4(\t\eta)} \sum_{\vk} \,
-|\chi'_k|^2+ \left(k^2-\frac{{\t{a}}'^2}{\t{a}^2} \right)
|\chi_k|^2+2 \frac{{\t{a}}'}{\t{a}} {\rm Re}( \chi_k \chi'^{\star}_k
)\, -\, C^{T}(k,\t\eta)\, , \ee
where the 4th adiabatic order subtraction terms are
\be C^{T}(k,\t\eta)= \frac{-{\t{a}}'^2+ \t{a} {\t{a}}''}{ 2\,
\t{a}^2 k}-\frac{6 {\t{a}}'^2 {\t{a}}''-3\t{a} {\t{a}}''^2-4 \t{a}
{\t{a}}' {\t{a}}^{'''}+{\t{a}}^2 {\t{a}}^{''''}}{8 {\t{a}}^3 k^3} \,
. \ee
Note that both vacuum expectation values, (\ref{renden}) and
(\ref{renpre}), are constant functions on $\man$ as they must be
given that the vacuum is translationally
invariant.%
\footnote{As a consequence, while the summands on the right side of
these expressions can be interpreted as the contribution of the mode
$\vk$ to $\langle \h{\rho}(x)\rangle_{\rm ren}$ and $\langle
\h{T}(x)\rangle_{\rm ren}$, they are \emph{not} the Fourier
transforms $\langle \h{\rho}_k \rangle_{\rm ren}$ and $\langle
\h{T}_k \rangle_{\rm ren}$ of these quantities in the $x$-space.}
If the vacuum were replaced by a generic 4th order adiabatic
state, this constancy will not hold. In that case the expectation
value would be distributions in $\x$. Finally, by themselves these
expectation values only provide a quadratic form on the Hilbert
space $\H_1$. However, recent results show that they are the
expectation values of an operator valued distribution $\h{T}_{ab}$
on $\H_1$ \cite{hollands}.

A natural question now arises: Does the vacuum expectation value of
the renormalized stress-energy tensor satisfy covariant conservation
$\t{\nabla}_a \langle \h{T}^{ab}\rangle_{\rm ren}=0$ with respect to
the dressed effective background $\t{g}_{ab}$? In any homogenous and
isotropic background, the conservation equation reduces to $\langle
\h \rho \rangle^\prime_{\rm ren} +\frac{\t{a}^\prime}{\t{a}} (4
\langle \h \rho\rangle_{\rm ren} + \langle \h T \rangle_{\rm
ren})=0$. The formal, unrenormalized expressions for $\langle \h
\rho \rangle$ and $\langle \h T \rangle$ satisfy this relation mode
by mode as a consequence of the wave equation satisfied by the
$\chi_k$. The adiabatic subtraction terms also satisfy the
conservation equation mode by mode. Thus, one can directly verify
that the expectation value of the renormalized stress-energy is
indeed conserved (see, for instance,
\cite{Anderson-Molina-Paris-Mottola}).

This adiabatic regularization of $\h{T}_{ab}$ also has the desired
properties enunciated in Wald's axioms \cite{waldbook}: it reduces
to the standard normal ordering in the flat space-time limit; the
subtractions terms are constructed from local information in the
background geometry, and, as already noted, the renormalized stress
energy is conserved; $\t{\nabla}_a \langle \h{T}^{ab} \rangle_{\rm
ren}=0$.

\emph{Remark:} As noted in section \ref{s3}, physically it is
appropriate to introduce an infrared cutoff by absorbing into the
background the modes whose physical wave length is larger than the
physical radius of the observable universe. Since physical wave
lengths scale linearly with the scale factor $\t{a}$, this can be
achieved in a time independent fashion by imposing a cut-off,
$k_{\rm IR}$, in the co-moving wave number $k$. As is clear from the
above discussion of conservation of stress-energy, the vacuum
expectation value of the new renormalized stress-energy tensor will
also be covariantly conserved. Furthermore, by construction, the
renormalized energy density and pressure will again be constant on
$\man$.\\

We will conclude this section with renormalization of the
Hamiltonian operator $\h{H}_1$ that generates the dynamics of
perturbations in conformal time $\t{\eta}$. The form
(\ref{pert-ham}) of the classical Hamiltonian and the fact that the
lapse corresponding to the conformal time is $N= a$ imply that the
Hamiltonian operator has the following formal expression:
\be \h{H}_{1,{\rm formal}} = \f{1}{2\Vzero}\,\, \sum_{\vk}
\f{4\kappa}{{\t{a}}^2}\, |\h{\p}_{\vk}|^2 + \f{{\t{a}}^2}{4\kappa}\,
k^2 |\h{\T}_{\vk}|^2  = {\t{a}}^4\int d^3x \, \h{\rho}_{\rm formal}\, .\ee
Therefore, the renormalized Hamiltonian is given by
\be \h{H}_{1,{\rm ren}} = \f{1}{2\Vzero}\,\, \sum_{\vk}
\f{4\kappa}{{\t{a}}^2}\, |\h{\p}_{\vk}|^2 + \f{{\t{a}}^2}{4\kappa}\,
k^2 |\h{\T}_{\vk}|^2 \,-\hbar \Vzero \, C^\rho(k,\t\eta)\, . \ee
Our discussion of energy density immediately implies that the
expectation values of $\h{H}_{1,{\rm ren}}$ are well defined on any
4th order adiabatic state. Furthermore, by taking its commutators
with $\h{\T}_{\vk}, \h{\p}_{\vk}$ we recover the Heisenberg
equations of motion (\ref{Teqn}) satisfied by $\h{\T}_{\vk}$.
However, we do not have a proof that $\h{H}_{1,{\rm ren}}$ is a
self-adjoint operator (in the precise sense that it is densely
defined and its domain equals that of its adjoint. In particular,
the strategy of using the Friedrich extension does not work because
the operator is not positive definite.) Although we expect this to
be the case, we also note that, quite generally in cosmology,
proving self-adjointness involves non-trivial subtleties and
technicalities
especially because the Hamiltonians have a non-trivial time dependence.\\

\emph{Remark:} In the case when $\man$ is topologically
$\mathbb{R}^3$, one continues to use the mode by mode subtraction
strategy but one has to replace $(1/\ell^3) \sum_{\vk}$ by
$1/(2\pi)^3 \int d^3 k$. Again, the renormalized stress-energy is
conserved, the subtraction terms are local and the prescription
reduces  to the standard normal ordering in Minkowski space-time.%
\footnote{Counter terms now appear under an integral sign. Results
of Ref. \cite{birrell,anderson-parker} imply that, one can use this
integral form of counter terms also in the spatially compact cases.
This procedure is justified on the grounds that the ultraviolet
regularization should not be sensitive to the global topology and
has, furthermore, some advantages.}
Again it is physically appropriate to introduce an infrared cut off
in the co-moving $k$-space and conservation of stress-energy
persists after imposing this cut-off. In addition, as shown in Ref.
\cite{birrell,anderson-parker}, in FRLW space-times this adiabatic
regularization is equivalent to the point-splitting Hadamard
renormalization. Therefore although the procedure is not manifestly
covariant because of the mode by mode subtraction, the result is
fully covariant w.r.t. $\t{g}_{ab}$. From a fundamental quantum
consideration, $\t{g}_{ab}$ is only a convenient mathematical
construction and the true quantum geometry is encoded in $\Psi_o$.
Still, it is desirable that the effective description have
space-time covariance.

\subsection{Narrowing down initial conditions}
\label{s6.3}

As discussed in section \ref{s6.1}, our quantum Hilbert space $\H_1$
of perturbations does not admit a preferred vacuum state. In quantum
field theories on Minkowski or de Sitter space-time, underlying
isometries serve as powerful tools to single out preferred states.
Our quantum FLRW background $\Psi_o$ is invariant under the
3-dimensional translation group acting on $\man = \mathbb{T}^3$. It
is then natural to seek preferred states in $\H_1$ by demanding that
they also be translationally invariant. As discussed in section
\ref{s5}, this condition leads us to the infinite dimensional space
of vacua which, in view of our discussion in \ref{s6.2}, we now
require to be of 4th adiabatic order. If one requires the state to
satisfy these properties initially, i.e. at the bounce, then they
are satisfied for all times. This is the family of `preferred'
states selected by the symmetry and regularity requirements. As we
will discuss in detail elsewhere, this choice is well-suited to
formulate a quantum version of Penrose's Weyl curvature hypothesis
\cite{penrose-weyl}. Furthermore, in the inflationary context, one
can motivate this choice using physical considerations based on
properties of the new repulsive force with origin in quantum
geometry that dominates the dynamics at and near the bounce
\cite{aan1,aan3}.

At a fundamental level, we allow any of these `vacua' as initial
conditions at the bounce provided, of course, the energy density
$\langle \h{\rho}\rangle$ in the perturbations at the bounce is
negligible compared to the energy density $\langle \h{\rho}_o
\rangle \sim 0.41 \rho_{\rm Pl}$ in the background quantum geometry
$\Psi_o$. However, for specific calculations and especially detailed
numerical simulations, one has to work with specific states. Are
there then especially convenient vacua to work with? The adiabatic
procedure summarized in section \ref{s6.1} provides a strategy to
select an `obvious' candidate, \emph{provided one fixes an instant
of time} $\t\eta_0$. Recall that to ensure 4th order adiabaticity we
required that, for $kL_4/a \gg 1$, the basis functions $\chi_k$ must
agree with a specific approximate solution $\chi_k^{(4)}$ defined in
(\ref{approxsol}) at least up to terms of adiabatic order four.
Therefore, given an instant $\t\eta_0$ of time, we can construct a
`natural' basis $\chi_k^{\rm obv}$ by asking that it has the same
initial data at that time as $\chi_k^{(4)}$. Thus the idea is to ask
for solutions $\chi_k^{\rm obv}(\t{\eta})$ to the exact evolution
equation (\ref{leq}) which satisfy
\be \label{obvious}  \chi_k^{\rm
obv}(\t\eta_0)=\chi_k^{(4)}(\t\eta_0)\, ; \quad \quad {\rm and}
\quad  \quad \partial_{\t\eta}\chi^{\rm
obv}_k(\t\eta_0)=\partial_{\t\eta} \chi^{(4)}_k (\t\eta_0)\, . \ee
Since $\chi_k^{(4)}$ are only approximate solutions to (\ref{leq}),
they will not agree with $\chi_k^{\rm obv}$ at any other time
$\t\eta\not=\t\eta_0$. Nonetheless, $\chi_k^{\rm obv}(\t\eta)$ are
automatically of 4th adiabatic order for all times $\t\eta$. We will
call the associated vacuum state the {\em obvious 4th-order
adiabatic vacuum at the time $\t\eta_0$.} In LQC, the preferred
instant of time required in this strategy is provided by the bounce.

Note, however, that for this strategy to work, the quantity
$W^{4}_k(\t{\eta}_0)$ in (\ref{approxsol}) must be non-negative
since it appears under a square-root in the expression of
$\chi_k^{(4)}(\t{\eta}_0)$. Now, as we pointed out in \ref{s3},
there is a natural \emph{physical} infrared cut-off $k_{\rm IR}$
provided by the radius of the observable universe. Typically, for $k
> k_{\rm IR}$, i.e, for the relevant $k$,\, $W_k^{(4)}$ is
positive. But if $k_{IR}$ is too low for this to happen, as
explained in section \ref{s6.1}, one has to suitably modify the form
of $W^{4}_k(\t{\eta}_0)$ for low $k$. In this case, even for a fixed
value $\t\eta_0$ of time, there would be an ambiguity in the choice
of the basis $\chi_k$ for low $k$, and hence also in the resulting
4th adiabatic order vacuum $|0\rangle^{\rm obv}_{{\t{\eta}}_0}$.

This vacuum state $|0\rangle^{\rm obv}_{{\t{\eta}}_0}$ is especially
convenient to work with in numerical simulations. Therefore it
serves as a technically powerful tool to establish the viability of
conjectures ---e.g., that, in the inflationary context, there exist
quantum states for which the back-reaction can be ignored--- and to
probe qualitative features of quantum dynamics. However, even when
there is a preferred instant of time, such as the bounce time in
LQC, and the infrared cut-off is sufficiently large for the state to
be unique, there is no physical principle that singles out this
vacuum over other 4th adiabatic order vacua. A better strategy would
be to narrow down the choice of vacua by adding external inputs
suggested by the physics of the problem. In the present analysis
there is a natural avenue along these lines. Since the test field
approximation plays a central role, one could ask: Are there 4th
adiabatic order vacua for which the expectation value of the
renormalized energy density is \emph{exactly} zero at a given time?
The answer to this question is in the affirmative and, although this
condition does not single out a unique state, it reduces the choices
considerably. This issue will be discussed in detail in a separate
publication \cite{zerostates}.

\subsection{ A Criterion for self-consistency of the truncation scheme}
\label{s6.4}

Our results in sections \ref{s5} -- \ref{s6.3} provide us with a
specific quantum theory, corresponding to the truncated phase space
$\t{\ps}_{\tr}$. This self contained mathematical framework enables
one to describe quantum dynamics of the truncated system, once the
initial state is specified, say at the bounce. \emph{The key
question now is whether the truncated theory has an interesting
domain of validity.} The basic assumption behind truncation is that
the back-reaction due to the stress-energy tensor $T_{ab}$ of the
first order fields can be neglected (compared to the background
stress-energy tensor $T^o_{ab}$) during the dynamical phase of
interest. Already in the classical theory, it is clear that generic
first order perturbations violate this condition. The interesting
question there is rather the following: Is there a sufficiently
large subspace in $\t\Gamma_{\tr}$ on which this condition is
satisfied? More precisely, if we restrict ourselves to an instant of
time and choose perturbations which do meet this condition, does the
condition continue to be satisfied over the
period of evolution of interest? %And if a perturbation meets this
%condition, is the it also met in an open neighborhood in
%$\t\Gamma^{(1)}$ of that perturbation?
We will now formulate the analogs of these questions in the quantum
theory.

Somewhat surprisingly, in the quantum theory one encounters some
novel features. First, in the classical theory the only `preferred'
time to specify the initial conditions would be the big bang which
is singular. In the quantum theory, by contrast, the bounce provides
a natural time $\t{\eta}_{\B}$ for this task. The second feature is
equally significant but also more subtle. Because of the underlying
symmetries of the background, it is natural to require the initial
state $\psi$ at time $\t{\eta}_{\B}$ to be translationally
invariant. An immediate consequence is that the expectation value
$\langle \h{T}_{ab} \rangle$ of the renormalized stress-energy
tensor operator is then homogeneous \emph{for all times}.%
\footnote{Note that this property holds even though the
perturbations operators $\h{\T}(\vec{x},\t\eta)$ themselves are
\emph{purely inhomogeneous}! In the classical theory, there are no
perturbations $\T(\vec{x}, \t{\eta})$ which are purely inhomogeneous
but their energy density is purely homogeneous.}
This property has the important consequence that the \emph{second
order} perturbations are again homogeneous and isotropic since their
sole source is $\langle \h{T}_{ab} \rangle$. Thus, thanks to the
symmetry of our initial state $\psi$ at the bounce, the back
reaction can change only the zeroth order, homogeneous fields,
changing the total state $\Psi_o \otimes \psi$ to a nearby state of
the type $(\Psi_o + \delta\Psi_o)\otimes \psi$.
% Now, the quantum states $\Psi_o$ we have chosen on the homogeneous
%sector are very sharply peaked on the effective trajectories. The
%back-reaction will shift these trajectories.
Our initial conditions on $\psi$ at $\t\eta= \t\eta_{\B}$ guarantee
that the shift $\delta\Psi_o$ at this initial time is negligible.
The key question then is whether it continues to remain negligible
under time evolution.

On general grounds one would say that the answer is dictated by the
time dependence of $\langle \h{T}_{ab} \rangle$. But in the detailed
mathematical framework, it is the second order truncation
$\h{\mathbb{S}}_2$ of the Hamiltonian constraint that determines the
change $\delta\Psi_o$ in the background quantum geometry and in this
equation it is only $\langle \h{\rho} \rangle$ ---rather than the
full $\langle \h{T}_{ab} \rangle$--- that enters as the source. Let
us suppose that the energy density $\langle \h{\rho} \rangle$ is
negligible compared to the background energy density
$\langle\h{\rho}_o\rangle$ from $\t\eta_{\B}$, until some time
$\t\eta_0$ of interest. Then it follows that $\delta\Psi_o$ would
continue to be negligible from $\t\eta{\B}$ to $\t\eta_0$. It may
seem somewhat surprising at first that one does not have to require
explicitly that the other independent component $\h{T}$ of
$\h{T}_{ab}$ should be small. This is because we are requiring that
$\langle \h{\rho} \rangle$ be negligible compared to $\langle
\h{\rho}_o \rangle$ not just initially but \emph{for all times
between $\t\eta_{\B}$ and $\t{\eta}_0$.} The difficult part of the
calculation is to check that the evolution of $\psi$ is such that
this condition does holds. If it does, then $\Psi_o\otimes \psi$
would provide a solution in which the back-reaction is negligible
even in the Planck era. This would then be a self-consistent
solution to the quantum truncated theory. \emph{Thus, a sufficient
condition for self-consistency of the truncation approximation is
that the energy tensity in the quantum perturbations should remain
small compared to that in the background from $\t\eta_{\B}$ to
$\t{\eta}_0$.}

In \cite{aan3} we will use detailed numerical simulations to show
that %in presence of a quadratic potential $V(\phi) = (1/2)
%m^2\phi^2$,
the inflationary scenario does admit states $\Psi_o\otimes \psi$ in
which $\langle \h{\rho}\rangle$ is negligible compared to the
background $\langle \h{\rho}_o \rangle$ for all times between the
bounce and the onset of the slow roll inflation. Furthermore, given
a state satisfying this condition, we will show that there is an
open neighborhood of $\psi$ such that the condition continues to be
satisfied by $\Psi_o\otimes \t\psi$ for all $\t\psi$ in this
neighborhood. Thus, there is a rich class of states that provide
self-consistent solutions to the truncated quantum theory,
demonstrating that the standard inflationary scenario admits a
consistent extension all the way back to the big bounce.

But the general framework constructed in this section is not tied to
inflation. It provides the technical machinery that is needed to
check if \emph{any} given paradigm, based on general relativity and
first order cosmological perturbations, admits a
\emph{self-consistent} extension to the Planck regime. More
precisely, it would enable one to address the following question: In
this paradigm, does the quantum theory admit solutions in which the
back reaction can be neglected throughout the period of interest,
including the Planck era?

\section{Summary and discussion}
\label{s7}

In the last four sections, we developed an extension of the standard
cosmological perturbation theory to include the Planck regime of
LQG. The strategy was to first truncate classical general relativity
coupled to a scalar field to the sector commonly used in the
cosmology of the early universe  ---FLRW space-times and linear
inhomogeneous perturbations thereon---  and then construct the
quantum theory of just this sector using LQG techniques.

Already in the truncation of the classical Hamiltonian theory there
is a subtlety that has often been overlooked in the LQC literature:
while the dynamics on the homogeneous sector is generated by a
(Hamiltonian) constraint, that on the full truncated phase space
$\ps_{\tr}$ is not. Indeed, because the dynamical vector field
$X^\alpha$ on $\ps_{\tr}$ fails to Lie drag the full symplectic
structure $\Omega_{\tr} = \Omega_o + \Omega_1$ on $\ps_{\tr}$, it is
not generated by \emph{any} Hamiltonian. Rather, $\ps_{\tr}$ is the
normal bundle over $\ps_o$ ---where the base space $\ps_o$ is
regarded as the homogeneous, isotropic subspace of the full phase
space $\ps$ of general relativity--- and $X^\alpha$ is the lift to
$\ps_{\tr}$ of the Hamiltonian vector field on $\ps_o$, induced by
the full Hamiltonian vector field on $\ps$. In the classical theory,
this subtlety can be ignored if one works with the space of
solutions (rather than the phase space), as is common in the
standard cosmology literature. But it becomes important for passage
to quantum theory if one wishes to treat both the perturbations and
the background quantum mechanically. Then there is no conceptual
justification for trying to construct dynamics for the \emph{full}
truncated system by imposing a quantum constraint. One can do this
only on the homogeneous sector, and one then has to `lift' this
quantum dynamics to the full Hilbert space just as in the classical
theory.

Having constructed the dynamics of gauge invariant variables on the
truncated phase space, we then used LQG techniques to construct
quantum kinematics: the Hilbert space $\H_o$ of states of background
quantum geometry, the Hilbert space $\H_1$ of gauge invariant
quantum fields $\h\Q,\, \h\T$ representing perturbations and
physically interesting operators on both these Hilbert spaces. The
imposition of the quantum constraint on the homogeneous sector leads
one to interpret the background scalar field $\phi$ as a
\emph{relational} or \emph{emergent} time variable with respect to
which physical degrees of freedom evolve. Furthermore, the
background geometry is now represented by a wave function $\Psi_o$
which encodes the probability amplitude for various FLRW geometries
to occur. The physically interesting wave functions $\Psi_o$ are
sharply peaked, \emph{but the peak follows a bouncing trajectory},
not a classical FLRW solution that originates at the big bang. In
addition, $\Psi_o$ has fluctuations about this bouncing trajectory.
Quantum fields $\h\Q,\, \h\T$, representing inhomogeneous scalar and
tensor perturbations, propagate on this \emph{quantum} geometry and
are therefore sensitive not only to the major departure from the
classical FLRW solutions in the Planck regime, but also to the
quantum fluctuations around the bouncing trajectory, encoded in
$\Psi_o$. Therefore at first the problem appears to be very
complicated. However, a key simplification made it tractable:
\emph{Within the test field approximation} inherent to the
truncation strategy, the propagation of $\h\Q,\, \h\T$  on the
quantum geometry $\Psi_o$ is completely equivalent to that of their
propagation on a specific, quantum corrected FLRW metric
$\t{g}_{ab}$. Although $\hbar$ does appear in its coefficients, this
`dressed, effective metric' $\t{g}_{ab}$ is smooth and allows us to
translate the evolution of $\h\Q,\, \h\T$ with respect to the
relational time to that in terms of the conformal (or proper) time
of $\t{g}_{ab}$. Furthermore, away from the Planck regime,
$\t{g}_{ab}$ satisfies Einstein's equations to an excellent
approximation. In this sense, the standard quantum field theory of
$\h\Q,\, \h\T$ emerges from the more fundamental description of
these fields evolving on the quantum geometry $\Psi_o$ with respect
to the relational time $\phi$. This exact relation between quantum
fields $\h\Q,\, \h\T$ on the quantum geometry $\Psi_o$ and those on
the dressed, effective geometry of $\t{g}_{ab}$ enabled us to carry
over adiabatic regularization techniques from quantum field theory
in curved space-times to those on quantum geometries $\Psi_o$.
Together, all this structure provides us with a well-defined quantum
theory of the truncated phase space we began with.

This framework has a broad range of applicability because scenarios
of the early universe are often based on linear perturbations on
FLRW backgrounds. Our construction provides an avenue to extend them
all the way to the quantum gravity era because quantum perturbations
now propagate on a quantum geometry which is completely regular,
with specific upper bounds for curvature and density in the
background. We can therefore use the new framework to re-examine the
`trans-Planckian issues'  encountered in these scenarios. Note first
that, the truncated theory under consideration here allows modes
with trans-Planckian frequencies. There is no obstruction because
the quantum geometry underlying LQG is subtle: In particular, while
there is a minimum non-zero eigenvalue of the area operator, there
is no such minimum for the volume or length operators even though
their eigenvalues are also discrete. \emph{The real danger is not
the existence of such modes but rather that the energy density in
these modes may not be negligible compared to that in the quantum
background geometry.} If this occurs, our quantum theory of the
truncated sector would not be viable. Whether this can happen is a
very non-trivial issue especially in the Planck regime immediately
following the bounce. Heuristically, if the state has just a few
excitations each carrying say, $10^6$ times the Planck energy in a
${\rm cm}^3$ volume, there would be no difficulty (since the energy
\emph{density} would be negligible). If on the other hand there is
one such excitation per Planck volume, our truncation approximation
will fail. Then we cannot neglect back-reaction. This is not an
impasse to quantum theory as such, but the proper treatment of such
states will have to await full LQG.

The key question then is whether the test field approximation
underlying this truncation scheme is satisfied. A priori this is a
difficult issue and, to our knowledge, had not been considered in
the literature because even to \emph{formulate} this question
precisely one needs the notion of the renormalized stress-energy
tensor on the Hilbert space $\H_1$ of quantum perturbations
propagating on the quantum geometry of $\Psi_o$. In our framework
this was provided by `lifting' the adiabatic techniques of
\cite{anderson-parker,parker-fulling74} to quantum fields on quantum
geometries $\Psi_o$. Specifically, by appealing to symmetry
principles and regularity requirements, we argued that it was
appropriate to focus attention on those states $\Psi_o\otimes \psi
\in \H_o\otimes \H_1$ of the combined system for which:

\indent 1) $\psi$ is invariant under the translational
symmetry of $\Psi_o$;\\
\indent 2) $\psi$ is a 4th order adiabatic state w.r.t.
$\t{g}_{ab}$; and, \\
\indent 3) \emph{at the bounce}, the energy density $\langle
\h{\rho}\rangle$ in the state $\psi$ is negligible\\
\indent\indent compared to the energy density $\langle \h{\rho}_o
\rangle$ in the background.

\noindent We then showed that the truncation approximation is
self-consistent if $\langle \h{\rho}\rangle$ continues to remain
negligible compared to $\langle \h{\rho}_o\rangle$ from the bounce
time $\t\eta_{\B}$ to a late time $\t{\eta}_0$ of physical interest
(e.g., when radiation decoupled from matter). In particular, our
argument shows that the full stress-energy tensor is not needed;
this significantly simplifies the task of performing numerical
simulations that are needed to check self-consistency.

As noted in section \ref{s1}, this criterion does not imply that
truncated solutions are necessarily close to exact solutions because
the sum of all higher order effects need not be negligible. However,
in practice such criteria are generally regarded as sufficient for
truncations to be trustworthy. Indeed, this philosophy governs the
entire theory of cosmological, stellar and black hole perturbations
in general relativity as well as perturbative calculations in
quantum field theory. In the same spirit, our self consistency
criterion can be used to test viability of the first order
truncation in the studies of the very early universe that include
the Planck regime.

In the next paper \cite{aan3} we will use this criterion in the
context of inflation. We first extend the general framework of this
paper slightly to incorporate the $(1/2) m^2\phi^2$ potential and
$\mathbb{R}^3$ spatial topology. (The value of $m$ is fixed by using
the 7 year WMAP data \cite{wmap,as3}.) We motivate the initial
conditions ---called `quantum homogeneity' at the bounce---  and
carry out detailed numerical simulations using the `obvious' 4th
order adiabatic vacuum for the initial quantum state $\psi$ of
perturbations. They show that the back-reaction of perturbations
remains negligible over the 11 orders of magnitude in matter density
and curvature, from the big bounce until the onset of slow roll.
Furthermore, the power spectrum at the end of inflation turns out to
be very close to that obtained in the standard inflationary scenario
and is thus compatible with the WMAP observations. By varying
initial conditions for the background (within computational
feasibility) we show that these results are robust. Furthermore, we
show that self-consistency is preserved if the initial state is
chosen to be in a neighborhood of the `obvious' 4th adiabatic order
vacuum. Taken together these results establish existence of
self-consistent extensions of the inflationary scenario to the
Planck regime. Finally, there is a small range for the value
$\phi_{\rm B}$ of the background inflaton field $\phi$ at the bounce
for which the quantum state at the onset of inflation differs
sufficiently from the Bunch Davies vacuum assumed in standard
inflation to give rise to non-Gaussianities that could be measured
in future observations along the lines of
\cite{agullo-parker,halo-bias1,halo-bias2,halo-bias3}.\\

We will conclude by pointing out a direction for significant
improvements and extensions of this framework. We began with a
truncation of general relativity coupled with matter, that is
well-suited for cosmology of the early universe. In the passage to
the quantum theory, for the homogeneous sector we used LQC
framework, rooted in the quantum geometry underlying LQG. This was
crucial for the resolution of the big bang singularity and the
subsequent quantum dynamics in the Planck era. On the other hand, to
facilitate comparisons with the standard cosmological literature, we
used a Fock-type representation for perturbations $\h\Q,\, \h\T$. As
in the Gowdy models \cite{hybrid1,hybrid2,hybrid3,hybrid4,hybrid5},
this is a well-defined and internally consistent quantization. But
it would be more satisfactory to use a `polymer-type' representation
rooted in LQG also for perturbations not only for aesthetic reasons
but also because it would provide sharper guidelines to relate the
truncated quantum theory to full LQG. Therefore, let us first ask:
Would this change of representation make a qualitative difference in
the results? The following heuristics lead us to believe that the
answer is in the negative. Note that $\h\Q,\, \h\T$ represent
\emph{perturbations} and, in any self-consistent solution
$\Psi_o\otimes\psi$, \emph{the energy density in the perturbations
is negligible compared to that in the background.}%
\footnote{This is a major conceptual difference from, say, the Gowdy
model where gravitational waves are not perturbations around any
`background' geometry. Indeed, in the vacuum Gowdy models, the
entire energy density resides in the gravitational waves.}
On general grounds one would expect that, in a viable LQG
representation of states capturing this physics, their dynamics and
properties would be well approximated by our $\psi\in \H_1$. As a
concrete illustration, we can use the following simplistic strategy
to aid intuition. Use the background structure available in the
truncated sector to decompose the perturbations $\Q(\x), \T(\x)$
into Fourier modes thereby representing these fields as an assembly
of harmonic oscillators, and then imagine using the polymer
representation of these oscillators to construct the LQG Hilbert
space $\H_1^{\rm LQG}$ for these fields. Then one knows
\cite{afw,cvz} that for low energy states the results of the polymer
and the standard (Fock-type) quantization are in excellent
agreement. This suggests the `hybrid' approach is viable from
practical or phenomenological perspective. Moreover, it is well
suited to bridge quantum field theory on quantum geometries with the
well-established quantum field theory in curved space-times.

However, from a fundamental perspective it is highly desirable to
systematically extend this framework by replacing $\H_1$ with an
appropriate Hilbert space that descends from LQG. In particular,
such an extension will enable one to arrive at the regularization
and renormalization procedure `starting from above' i.e., from full
LQG considerations. By contrast, in this paper, we have introduced
this procedure `starting from below', i.e., from quantum field
theory in curved space-times. Put differently, our primary goal of
this paper is to carve out a path to extend the cosmological
perturbation theory to the Planck era. The emphasis has been on
showing that there does exist such a framework with a number of
desirable mathematical properties and physical features which,
moreover, is well suited for phenomenological applications. But from
a fundamental LQG perspective, it can and should be related to LQG
even more closely. %further improved.

\section*{Acknowledgments}

We would like to thank David Brizuela, Stefan Hollands, Alok Ladha,
Jerzy Lewandowski, Hanno Sahlmann and Thomas Thiemann for
discussions and Guillermo Mena, Parampreet Singh, David Sloan and
Jose Velhinho for correspondence. This work was supported in part by
the NSF grant PHY-1205388, the Eberly research funds of Penn state
and the Marie Curie Fellowship program of the EU.

\begin{appendix}

\section{Truncated dynamics: A simple example}
\label{a1}

In this appendix we will illustrate the truncation procedure  of
section \ref{s3} using $\lambda \Phi^4$-theory. This example is
simple enough to perform explicit calculations that bring out the
main conceptual subtleties ---in particular the differences between
dynamics of the exact and truncated theories--- which are sometimes
overlooked in the cosmology literature.

\subsection{Space-time framework}
\label{a1.1}

As in the main text we assume that the space-time $M$ is
topologically $\man\times \mathbb{R}$, where the Cauchy surfaces
$\man$ are topologically 3-toruses $\mathbb{T}^3$. But the
space-time metric $\eta_{ab}$ is now assumed to be flat (with
signature -,+,+,+). Denote by $\S$ the space of suitably regular
solutions to
\be \label{kg} \Box \Phi -\mu^2\Phi -\lambda\Phi^3 =0\ee
and by $\S_o$ its subspace consisting of spatially homogeneous
solutions. We are interested in a small neighborhood of $\S_o$ in
$\S$. In this neighborhood, it is convenient to consider curves
\be \label{expansion} \Phi[\ep](\x,t) = \phi(t) + \ep \vpone(\x,t) +
\f{\ep^2}{2!}\, \vptwo(\x,t) + \ldots + \f{\ep^n}{n!} \vpn(\x,t) +
\dots\ee
parameterized by $\ep \in ]-1,\, 1[$, which pass through $\S_o$ at
$\ep=0$. The $\vpn(\x,t)$ are to be thought of as the $n$th order,
inhomogeneous perturbations on the homogeneous solution $\phi(t)$.
Since we are interested in curves that move away from $\S_o$, to
avoid redundancy, without any loss of generality we will assume that
the first order perturbation $\vpone(\x,t)$ are \emph{purely}
inhomogeneous, i.e. that
\be \sint d\vzero\, \vpone (\x,t) = 0 \qquad \forall t\, . \ee
Here in what follows all the integrals are over $\man$ and $d\vzero$
is the natural volume element thereon. Substituting the expansion
(\ref{expansion}) in (\ref{kg}) and matching coefficients of $\ep^n$
for each $n$ we obtain a hierarchy of equations:
\ba  \ddot{\phi} + \mu^2\phi +\lambda \phi^3  = 0, &\qquad&  (\Box
- \mu^2 -3\lambda \phi^2) \vpone = 0, \nonumber\\
(\Box - \mu^2 - 3\lambda \phi^2) \vptwo = 6\lambda \phi (\vpone)^2,
&\quad& (\Box - \mu^2 - 3\lambda \phi^2)\, \vpthree = 6\lambda\,
((\vpone)^3 + 3 \phi \vpone\vptwo), \nonumber\\
\label{hierarchy}\ldots &\qquad& \ldots \ea

Note that the first equation on $\phi$ is non-linear but an
\emph{ordinary differential equation} (ODE) and each of the
subsequent equation on $\vpn$ is a \emph{linear} partial
differential equation (PDE) in $\vpn$ with sources containing lower
order fields, already determined by solving the previous equations
in the hierarchy. Thus, the task of solving the non-linear PDE
(\ref{kg}) is reduced to solving one non-linear ODE and a succession
of linear PDEs.%
\footnote{Note that the equation on $\vpone$ has no source terms.
Therefore, if the initial data is purely inhomogeneous, so is the
solution. Because equations of $\vpn$ for $n>1$ have source terms
which are non-linear in the lower order fields, we cannot demand
that they be purely inhomogeneous. Note also that, for $n>1$ there
is freedom in adding a solution to the homogeneous equation. This
can be removed, e.g., by choosing retarded solutions to the
inhomogeneous equations.} %The idea is that if the truncation of the sum in
%(\ref{expansion}) to $n$th order has the same initial data as that
%of $\Phi$ of (\ref{kg}) in the infinite past, the truncated solution
%would approximate $\Phi$.}
%
The idea is that an approximate solution to the full problem can be
obtained by truncating the series to the appropriate order, i.e., by
ignoring terms of $O(\ep^{n+1})$, say. As usual, while the meaning
of the approximation in terms of the smallness parameter $\epsilon$
(which could be tied to the coupling constant $\lambda$ for physical
reasons) is clear, the truncated series can be a good approximation
to the full solution $\Phi$ only if $\vpn$ remain small compared to
$\phi$. \emph{It is important to note that this scheme of obtaining
approximate solutions is distinct from an alternative procedure that
appears to be used often in the cosmology literature} (although
sometimes only implicitly). That strategy corresponds to defining a
field $\delta\Phi$ via $\Phi(\x,t) =\phi(t) +\delta\Phi$ and solving
the equation
\be \Box \delta\Phi  -\mu^2 \delta\Phi - (3\lambda \phi^2)
\delta\Phi - (3\lambda\phi)\, \delta\Phi^2 -\lambda \delta\Phi^3
=0\ee
To the linear order this strategy agrees with (\ref{hierarchy}), but
not to higher. Indeed, already at the second order, one now has to
solve a \emph{non-linear} PDE, with a function $\phi$ as a
coefficient, which is in some ways more complicated than solving the
original (\ref{kg}).

\subsection{The Hamiltonian framework}
\label{a1.2}

For the $\lambda\Phi^4$ system, one could pass to the quantum theory
directly from the classical space-time formulation sketched above.
However, general relativity is a background independent theory and
the generalized Dirac quantization strategy followed in LQC requires
us to pass through the Hamiltonian framework. Therefore we will now
illustrate how the truncation procedure of section \ref{a1.1} works
in the phase space language.

The full phase space $\ps$ is spanned by pairs $(\Phi(\x), \Pi(\x))$
on the 3-manifold $\man$ which is topologically $\mathbb{T}^3$ and
its homogeneous subspace $\ps_{o}$ is spanned by real numbers
$(\phi, \pphi)$. The symplectic structure is given by:
\be \Omega (\delta_1,\delta_ 2) = \sint d\vzero \,\,\,
[\delta_1\Phi\, \delta_2\Pi - \delta_2\Phi\, \delta_1\Pi]\, , \ee
where $\delta \equiv (\delta\Phi,\, \delta\Pi)$ denotes tangent
vectors to $\ps$. The corresponding Poisson brackets are also the
familiar ones: $\{\Phi(\x_1), \,\, \Pi(\x_2)\} = \delta
(\x_1,\x_2)$. Dynamics is generated by the Hamiltonian $H$:
\be H(\Phi,\Pi) = \f{1}{2}\, \sint d\vzero\, [\Pi^2 + D_a\Phi
D^a\Phi + \mu^2 \Phi^2 + \f{\lambda}{2} \, \Phi^4]\, .   \ee
Note that although $\ps$ is infinite dimensional just as the
solution space $\S$, and $\ps_o$ is 2-dimensional just as $\S_o$,
there is a key difference: $\ps$ and $\ps_o$ are vector spaces
unlike $\S$ and $\S_o$. As we will see the ability to define
constant vector fields on $\ps$ plays an important role in defining
truncated dynamics. (This structure is also available on the phase
space of general relativity and used in the main text to define
truncated dynamics.)

Again, we are interested only in a small neighborhood of the
homogeneous subspace $\ps_o$ of $\ps$. Therefore we are led to
consider 1-parameter family of curves, parameterized by $\ep\in
]-1,1[$ :
\ba \Phi[\ep](\x) &=& \phi + \ep \vpone (\x) +\, \ldots\,
+\f{\ep^n}{n!} \vpn (\x) +\,\ldots\quad {\rm and} \nonumber\\
\Pi[\ep](\x) &=& \f{\pphi}{V_o} + \ep \pione (\x) + \,\ldots \,
+\f{\ep^n}{n!} \pin (\x) +\, \ldots\ea
where $\vpone(\x), \pione(\x)$ are purely inhomogeneous, and $V_o$
is the volume of the 3-torus $\man$. By truncating the series at
$n$th order we obtain the phase space that is appropriate for
describing the homogeneous solutions $(\phi, \pphi)$, together with
first, second,\, \ldots\, $n$th order perturbations propagating
thereon. Since in the main text we focus on just the first order
perturbations, let us do the same here.

Thus, let the truncated phase-space $\ps_{\tr}$ consist of a doublet
of pairs of canonically conjugate fields $(\phi,\pphi;\,
\vpone(\x),\pione(\x))$ where $\phi,\pphi$ are real numbers
(representing homogeneous fields on $\man$) and $\vpone(\x),
\pione(\x)$ are purely inhomogeneous fields on $\man$. Thus,
$\ps_{\tr} = \ps_o \times \ps_1$. The symplectic structure on
$\ps_{\tr}$ is given by $\Omega_{\tr} = \Omega_o + \Omega_1$, where
\be \Omega_o(\delta_1, \delta_2) = \delta_1\phi\, \delta_2\pphi -
\delta_2\phi\, \delta_1\pphi\, \qquad \Omega_1(\delta_1, \delta_2) =
\sint d\vzero \, [\delta_1\vpone \delta_2\pione - \delta_2\vpone
\delta_1\pione] \ee
which yields the Poisson brackets
\be \{\phi,\, \pphi\} = 1, \qquad  \{\vpone(\x_1),\, \pione(\x_2)\}
= [\delta (\x_1,\x_2) - \f{1}{V_o}]  \equiv
\bar{\delta}(\x_1,\x_2)\ee
where $\bar\delta(\x_1,\x_2)$ is the Dirac delta-distribution
restricted to the purely inhomogeneous fields on $\man$.%
\footnote{The term $1/V_o$ is necessary simply because the fields
$\vpone(\x)$ and $\pione(\x)$ are purely inhomogeneous. For example,
if we integrate the left side of the Poisson bracket between
$\vpone$ and $\pione$ over $x_1$ (or $x_2$), we get zero and $1/V_o$
terms assures that the right side also vanishes.}
By decomposing fields $\Phi(\x), \Pi(\x)$ in $\ps$ into purely
homogeneous and purely inhomogeneous parts, one can readily see that
the full pase space $(\ps,\, \Omega)$ is naturally isomorphic to the
truncated phase space $(\ps_{\tr}, \Omega_{\tr})$. However, the
physical meaning of the inhomogeneous fields is different in the two
cases and, more importantly, the dynamics is very different.

Geometrically, $\ps_{\tr}$ is the normal bundle over $\ps_o$ (since
purely inhomogeneous fields are orthogonal to the purely homogeneous
ones in the $L^2$ norm of the space of functions on $\man$.) To
obtain the dynamical flow on $\ps_{\tr}$, let us begin with the
full, non-linear Hamiltonian vector field $X_H$ on a small
neighborhood of $\ps_o$ in $\ps$. Since the exact Hamiltonian flow
generated by $X_H$ is tangential to $\ps_o$, the equations of motion
on $\ps_o$ are just the restrictions of those on full $\ps$ to the
homogeneous sector: $\dot\phi = \pphi/V_o, \qquad \dot{p}_{(\phi)} =
-(\mu^2\phi +\lambda \phi^3)V_o$. To obtain the equations of motion
on the (inhomogeneous) tangent vectors $(\vpone,\pione)$, we first
note that the Hamiltonian flow $X_H$ on $\Gamma$ naturally drags
these tangent vectors along dynamical trajectories on $\Gamma_o$. To
obtain the `dot', however, we need to compare the image $(\vpone,
\pione)|_{t+ \delta t}$ at $(\phi,\pphi)|_{t + \delta t}$ with the
original tangent vector $(\vpone, \pione)|_t$ at the point
$(\phi,\pphi)|_t$ of $\ps_o$. This can be trivially accomplished
because $\ps$ has a vector space structure. The resulting equations
of motion are: $\dot{\varphi}^{(1)}(\x,t) = \pione(\x,t), \,\,
\dot{\pi}^{(1)}(\x,t) = [D^2  -\mu^2 -3\lambda \phi^2(t) ]\vpone
(\x,t)$. The resulting dynamical vector field $X_{\dyn}$ on
$\ps_{\tr}$ is given by:
\ba X_{\dyn} &:=& (\dot\phi, \dot{p}_{(\phi)};\,
\dot{\vp}^{(1)},\dot{\pi}^{(1)}\nonumber\\
&=& \, \Big(\,\f{\pphi(t)}{V_o},\, -[\mu^2\phi(t) +\lambda
\phi^3(t)]V_o;\,\,\, \pione,\, [D^2 -\mu^2 -3\lambda \phi^2(t)
]\vpone\, \Big)\nonumber\\ \ea

Thus, the full dynamics on $\ps$ induces a well-defined flow
$X_{\dyn}$ on $\ps_{\tr}$. Furthermore, this dynamical vector field
can be expressed in a form adapted to symplectic geometry. Using
Greek letters to denote the abstract indices labeling tangent
vectors to $\ps$,\,\, $X^\alpha_{\dyn}$ can be expressed as
\be X_{\dyn}^\alpha = \Omega^{\alpha\beta}_o
\partial_\beta H_o + \Omega^{\alpha\beta}_1 \partial_\beta H_1\ee
where
\ba H_o(\phi,\pphi) &:=& H|_{\ps_o} = \f{V_o}{2}\,\,
\Big(\f{\pphi^2}{V_o^2} +
\mu^2\phi^2 + \f{\lambda}{2})\phi^4 \Big), \nonumber\\
H_1 (\vpone,\pione) &=& \f{1}{2}\, \sint d\vzero\, [(\pione)^2 +\,
D^a\vpone\, D_a \vpone\, + \mu^2 (\vpone)^2 +
3\lambda(\phi^2)(\vpone)^2]\, . \ea
Note however that this is \emph{not} a Hamiltonian flow on
$\ps_{\tr}$ because it is not of the form $(\Omega_o^{\alpha\beta} +
\Omega_1^{\alpha\beta})\partial_\beta H_{\tr}$ for any function
$H_{\tr}$ on $\ps_{\tr}$. The obvious candidate, $H_{\tr} := H_o +
H_1$ does not work because $\Omega^{\alpha\beta}_o \partial_\beta
H_1 \not= 0$ (since $H_1$ depends not only on $(\vpone,\pione)$ but
also on $\phi$).%
\footnote{Had we simply defined dynamics using $H_{\tr} := H_o
+H_1$, it would have included the homogeneous part of the back
reaction of the first order perturbation $(\vpone, \pione)$. This
would not be physically consistent because this dynamics ignores the
inhomogeneous part of the back-reaction of the same perturbation
which is of the same order in the $\ep$ expansion.}
More generally, one can verify that the fact that $H_1$ depends on
$\phi$ implies that the Lie derivative of $\Omega_{\tr} := \Omega_o
+ \Omega_1$ by the dynamical vector field $X_{\dyn}$ on $\ps_{\tr}$
does not vanish.

In the classical theory, the fact that we have a well-defined
dynamical flow on $\ps_{\tr}$ suffices. However, the fact that the
flow is not Hamiltonian introduces new features in the transition to
quantum theory. Keeping the quantum perspective in mind, one can
rephrase the classical dynamics as follows. Since the homogeneous
solution is to be regarded as the background and inhomogeneities as
perturbations, we can first restrict ourselves to the homogeneous
part of the phase space $(\ps_o, \Omega_o)$ and note that, on it,
the dynamics is indeed governed by a true Hamiltonian flow,
generated by $H_o$. Fix any dynamical trajectory $\phi(t), \pphi(t)$
in $\ps_o$. To specify how perturbations propagate on this
background solution, we need to lift this trajectory to the normal
bundle, $\ps_{\tr}$. This is precisely what the remaining part,
$\Omega_1^{\alpha\beta}\partial_{\beta}\, H_1$, of the dynamical
vector field $X^\alpha_{\dyn}$ does. Given any tangent vector
$(\vpone_o,\,\pione_o)$ at a point, say $(\phi(t_o), \pphi(t_o))$,
along the given dynamical trajectory, the orbit of $X^\alpha_{\dyn}$
through the point $(\phi(t_o), \pphi(t_o);\, \vpone_o(t_o),
\pione_o(t_o))$ of $\ps_{\tr}$ specifies the dynamics of the
perturbation $(\vpone,\,\pione)$ on the background trajectory
$(\phi(t), \pphi(t))$. The fact that we have lifts to $\ps_{\tr}$ of
orbits in $\ps_o$ trivially implies that, even though different
choices of the initial $(\vpone,\pione)$ define distinct orbits
(describing dynamics of various perturbations), all these orbits in
$\ps_{\tr}$ project down to the same orbit on $\ps_o$. This is just
a reflection of the fact that the dynamics defined by
$X^\alpha_{\dyn}$ neglects the back-reaction of the perturbation on
the homogeneous background.\medskip

\emph{Remark:} To obtain the truncated dynamics $X_{\tr}$ we used
the vector space structure of $\ps$ to transport the vector
$(\vpone,\pione)|_{t+\delta t}$ at $(\phi, \pphi)|_{t + \delta t}$
to the point $(\phi, \pphi)|_{t}$. However, one can carry out this
comparison more generally, e.g., if there is a natural flat
connection to transport vectors from one point of the phase space to
another. This is the case when $\ps$ is a cotangent bundle over an
affine space (as in LQG) or over a convex subset of a vector space
(as in the ADM framework of general relativity). Therefore the
procedure of inducing dynamics on the truncated phase space from the
exact dynamics in a neighborhood of the homogeneous sector of the
phase space goes through also in the cosmological context analyzed
in the main text.

%\section{Inclusion of a potential $V(\phi)$ in cosmology}%
%\label{a2}

\end{appendix}

\end{document}